\documentclass[aps, pre, amsfonts, amssymb, amsmath, twocolumn, notitlepage, nofootinbib, twoside]{revtex4-2}

\usepackage{graphicx}
\usepackage{dcolumn}
\usepackage{gensymb}
\usepackage{color}
\usepackage[dvipsnames]{xcolor}
\usepackage{layout}
\usepackage[english]{varioref}
\usepackage[greek,english]{babel}
\usepackage{subfigure}
\usepackage[mathcal]{eucal}
\usepackage{xfrac}
\usepackage{multirow}
\usepackage{listings}
\usepackage{bm}
\usepackage{comment}
\usepackage[normalem]{ulem}
\usepackage[babel]{microtype}
\usepackage{xr-hyper}
\usepackage[pdftex, pdftitle={Article}, pdfauthor={Author}, colorlinks=true,linkcolor=blue,citecolor=blue,urlcolor=blue]{hyperref}

\newcommand{\AB}[1]{\textcolor{blue}{[AB: #1]}}

\begin{document}

\title{Anisotropic DLVO-like interaction for charge patchiness in colloids and proteins}

\author{Andra{\v z} Gnidovec}
\affiliation{Faculty  of  Mathematics  and  Physics,  University  of  Ljubljana,  Ljubljana,  Slovenia}
\author{Emanuele Locatelli}
\affiliation{Department of Physics and Astronomy, University of Padova, 35131 Padova, Italy} 
\affiliation{INFN, Sezione di Padova, via Marzolo 8, I-35131 Padova, Italy}
\author{Simon {\v C}opar}
\affiliation{Faculty  of  Mathematics  and  Physics,  University  of  Ljubljana,  Ljubljana,  Slovenia\relax}
\author{An{\v z}e Bo{\v z}i{\v c}}
\affiliation{Department of Theoretical Physics, Jo\v{z}ef Stefan Institute, Jamova 39, SI-1000 Ljubljana, Slovenia}
\author{Emanuela Bianchi}
\email{emanuela.bianchi@tuwien.ac.at}
\affiliation{Institut f{\"u}r Theoretische Physik, TU Wien, Wiedner Hauptstra{\ss}e 8-10, A-1040 Wien, Austria} 
\affiliation{CNR-ISC, Uos Sapienza, Piazzale A. Moro 2, 00185 Roma, Italy}

\begin{abstract} 
The behaviour and stability of soft and biological matter depend significantly on electrostatic interactions, as particles such as proteins and colloids acquire a charge when dispersed in an electrolytic solution. A typical simplification used to understand bulk phenomena involving electrostatic interactions is the isotropy of the charge on the particles. However, whether arising naturally or by synthesis, charge distributions are often inhomogeneous, leading to an intricate particle-particle interaction landscape and complex assembly phenomena. The fundamental complexity of these interactions gives rise to models based on distinct assumptions and varying degrees of simplifications which can blur the line between genuine physical behaviour and artefacts arising from the choice of a particular electrostatic model. Building upon the widely-used linearized Poisson-Boltzmann theory, we propose a theoretical framework that -- by bridging different models -- provides a robust DLVO-like description of electrostatic interactions between inhomogeneously charged particles. By matching solely the {\em single-particle} properties of two different mean-field models, we find a quantitative agreement between the {\em pair interaction energies} over a wide range of system parameters. Our work identifies a strategy to merge different models of inhomogeneously charged particles and paves the way to a reliable, accurate, and computationally affordable description of their interactions.
\end{abstract}

\maketitle


Electrostatics is omnipresent in soft matter and biology since colloids and proteins dispersed in an electrolytic medium are inevitably charged due to dissociation of ionizable groups and/or preferential adsorption of ions from the surrounding solvent~\cite{yang2023jpcb,hueckel2021total,zhou2018electrostatic}. On colloidal surfaces, these processes typically lead to charged particles with charge inhomogeneities that are negligible with respect to particle size ($100$~nm--$1\,\mu$m). However, they have been recently exploited to synthesize colloids with charged surface patterns engineered to drive material assembly~\cite{vanostrum2015jpcm,sabapathy2017pccp,Zimmermann_2018,kierulf2022starch,Virk2023jpcm}. In contrast, proteins are intrinsically charged in a highly uneven fashion as the (de)protonation of their ionizable amino acids leads to charge inhomogeneities comparable in scale to the protein size ($\sim$1 nm). The resulting surface charge distribution of a protein is then dictated by the configuration of acidic and basic groups along its backbone~\cite{zhou2018electrostatic}.

Charge inhomogeneity -- or {\em patchiness} -- has a significant impact on the behaviour of charged systems, as it is not only highly directional~\cite{Persson2015jpcb,kress2020colloidal} but can induce attraction between neutral and even like-charged objects~\cite{adar2017electrostatics,lunkad2022both}. Control over charge patchiness can be achieved both directly {\em via} targeted synthesis of colloids or mutagenesis of proteins and indirectly by inducing changes in the solvent pH~\cite{zhou2018electrostatic,kim2021patchy}. This makes charge patchiness a convenient parameter for tuning the properties of electrostatic interactions. Despite the synthesis of charged patchy colloids with simple surface patterns still being in its infancy, the experiments conducted thus far highlight the fact that the anisotropic nature of interparticle interactions can guide the formation of aggregates with nontrivial, non-close-packed architectures~\cite{Kimura_2019,lebdioua2021jcis,naderi2020self,li2022modulating}. In protein systems, the intrinsic inhomogeneity of surface charge has recently emerged as a powerful feature that influences protein aggregation~\cite{guo2021quantifying,zhang2020assembly} and liquid-liquid phase separation~\cite{zhou2018electrostatic, Knowles_2023,kim2024surface}.

When describing electrostatic interactions between inhomogeneously charged particles, the level of detail is usually adjusted to the phenomenon under consideration. Analytical approaches, which typically rely on the linearized Poisson-Boltzmann (PB) approximation, have a long history (see Refs.~\cite{besley2023recent,siryk2021jcp} for an overview), with the recent decade bringing notable advances in analytical solutions for the interactions between inhomogeneously charged particles~\cite{Boon2012,bozic2013jcp,siryk2021jcp,obolensky2021rigorous,bianchi2011sm,bianchi:2015,Luijten2016}. These are, however, difficult to efficiently evaluate when particles carry complicated patterns of charge and become prohibitively expensive when applied to large-scale systems. Models of charged patchy particles consequently rely on simplifications to make them computationally feasible~\cite{bianchi2011sm,bianchi:2015,popov2023jpcb,yigit15a} and enable large-scale simulations of their assembly in two and three dimensions~\cite{mani2021stabilizing,silvanonanoscale,Swan_2019,yigit17}. Yet the same simplifications often make it unclear which assembly properties are a true consequence of patchy particle interactions and which properties are an artefact of a given model.

We develop a unifying analytical framework that connects mean-field descriptions of inhomogeneously charged particles at different levels of approximation. Within this framework, we obtain an excellent agreement for {\em pairwise interactions} predicted by two distinct electrostatic models~\cite{bianchi2011sm,bianchi:2015,bozic2013jcp} across a wide range of system parameters, which is achieved solely by matching their {\em single-particle} properties. This approach allows better control over the approximations used in the models, resulting in a reliable and easy-to-implement description of interactions between inhomogeneously charged particles. As such, it works as an anisotropic generalization of the electrostatic component of the DLVO theory~\cite{siryk2021jcp,VerweyOverbeek-1948}.

\subsection*{Reconsidering existing models of charged patchy particles}

The two models on which we build our theoretical framework describe electrostatic interactions between inhomogeneously charged spherical particles in the linearized Debye-Hückel (DH) regime. To distinguish between the two models, we will term them IC (``internal charge'') and CS (``charged shell'') and use a suffix to describe whether or not the ions from the solvent outside the particles are present also inside them (suffix ``i'' standing for \emph{impermeable} and suffix ``p'' for \emph{permeable} particles). By rigorously relating these two models to each other, we can develop a unified mean-field description of interactions between charged patchy particles which also allows for a further refinement of the two models.

\begin{figure*}[!ht]
\centering  
\includegraphics[width=\textwidth]{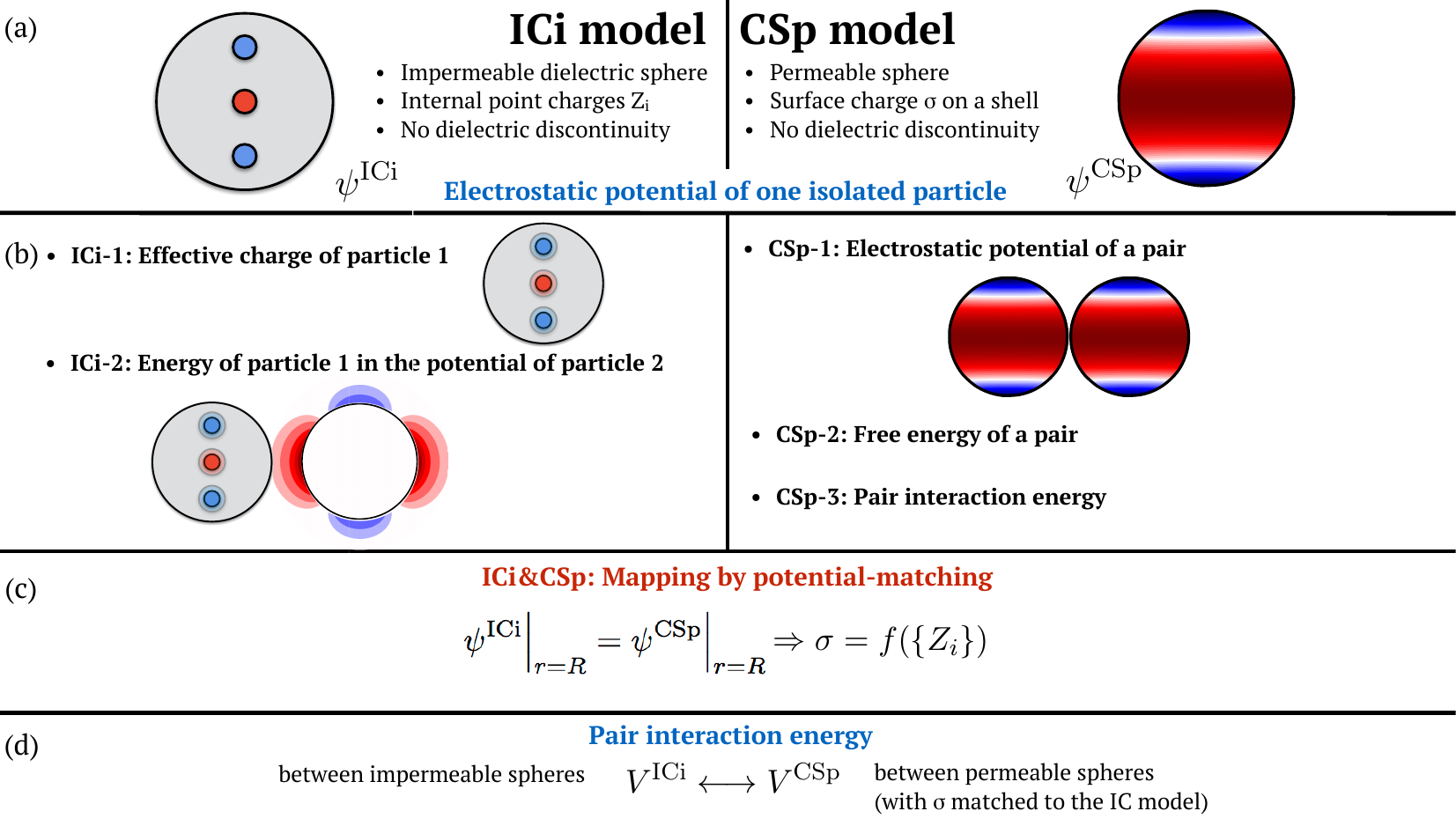}
\caption{Sketch of the proposed theoretical framework to determine the electrostatic interactions between inhomogeneously charged particles. Detailed formulation is given in Methods and SI. {\bf (a)} Essential properties of impermeable internal charge (ICi) and permeable charged shell (CSp) models. {\bf (b)} Basic steps in the determination of the pair interaction energies within the two models. {\bf (c)} Mapping between the two models achieved by redefining the surface charge distribution of the CSp model so that the surface potential of a {\em single} CSp particle matches the one of an ICi particle. {\bf (d)} Comparison of the resulting electrostatic pair interaction energies of the potential-matched models.}
\label{fig:approach}
\end{figure*}

The IC model was introduced by Bianchi {\em et al.}~\cite{bianchi2011sm,bianchi:2015} to describe interaction patterns of colloids synthesized with differently charged surface regions~\cite{vanostrum2015jpcm,sabapathy2017pccp,Zimmermann_2018,kierulf2022starch,Virk2023jpcm}. These inhomogeneously charged colloids are typically ion-impenetrable and thus modelled as impermeable dielectric spheres carrying a discrete number of point charges $\{Z_i\}$ in their interior (ICi; Fig.~\ref{fig:approach}a, left). The number and positions of the point charges are chosen in such a way that they qualitatively reproduce the symmetries of the charge pattern on the particle surface. The effective pair interaction energy in the ICi model is determined by first calculating the electrostatic potential around a single particle and then approximating the pairwise interaction as the energy of a ``probe'' particle in the electrostatic potential of the ``source'' particle. In the interior of the probe particle, the original charges must be dressed with an excluded volume factor that takes into account that the $\{Z_i\}$ are embedded in an impermeable sphere and thus cannot be approached by solution ions (Methods and Fig.~\ref{fig:approach}b, left).
In the original model by Bianchi {\em et al.}~\cite{bianchi2011sm,bianchi:2015} the effective charge was approximated using the excluded volume factor of the DLVO theory~\cite{siryk2021jcp,VerweyOverbeek-1948}, i.e., $Z^*_i= \exp(\kappa R)/(1+\kappa R)Z_i$ for all charges $\{Z_i\}$. We improve on this approximation by determining the effective charges in a general fashion. Specifically, we require that the surface electrostatic potential generated by a permeable particle containing effective point charges matches the one generated by the point charges of its impermeable counterpart. Notably, this is done at the {\em single-particle} level by matching the leading terms of the multipole expansion of the electrostatic potentials of the two particles (Methods and Supporting Information (SI)).

The CS model was introduced by Bo\v{z}i\v{c} and Podgornik with the aim to describe the electrostatic interactions between virus particles which typically posses icosahedrally symmetrical inhomogeneous charge distributions~\cite{bozic2012jbp}. As virus capsids are permeable to solution ions, this model features inhomogeneously charged particles as permeable charged spherical shells with a continuous surface charge distribution $\sigma$ (CSp; Fig.~\ref{fig:approach}a, right). The CSp model allows for an analytical derivation of the exact {\em pair} electrostatic potential of the two-particle system, expressing the free energy of pair interaction as the work associated with moving ions from the bulk of the solution to the particle surfaces (Methods and Fig.~\ref{fig:approach}b, right).

On the one hand, the IC model allows for a fast numerical evaluation of pair energies but involves several simplifications and cannot reproduce any arbitrary surface charge distribution. On the other hand, the ability of the CS model to reproduce a wide array of surface charge distributions comes at the expense of a larger computational demand. Another important difference between the two models lies in the fact that ICi particles are ion-impenetrable while CSp particles are ion-penetrable.

\subsection*{Unifying different electrostatic models by matching single-particle potentials}

In order to unify the two electrostatic models into a single framework, we define an {\em analytical} procedure that maps discrete charges of the ICi model to a continuous surface charge distribution of the CSp model and bridges the gap between impermeable and permeable particles. The mapping procedure is based on {\em single-particle} properties of the two descriptions with the aim to extend this match to their {\em pair interaction energies}. This approach is motivated by the recent work of \citet{everts2020screened}, who has shown that such a mapping in the other direction -- from impermeable bodies with a continuous surface charge density to permeable bodies with point, line, or surface charges -- while not unique, can in specific cases guarantee the preservation of the pair interaction energy.

Since particle radius $R$, dielectric constant of the solvent $\varepsilon_w$, and Debye screening length $\kappa^{-1}$ are all shared between the two models by construction, the only particle feature that remains to be defined via this mapping is the particle charge. Note that, for sake of simplicity, we assume 
there is no dielectric jump. To obtain the relation between the ensemble of the discrete charges of the ICi model, $\{Z_i\}$, and the surface charge distribution of the CSp model, $\sigma$, we require that the single-particle electrostatic potential $\psi^\mathrm{ICi}$ at the surface of an impermeable particle must be identical to the single-particle electrostatic potential $\psi^\mathrm{CSp}$ at the surface of its permeable counterpart (Fig.~\ref{fig:approach}c). This can be synthetically expressed as
\begin{equation}
\psi^\mathrm{ICi} \Big|_{r=R} = \psi^\mathrm{CSp}\Big|_{r=R}  \Rightarrow \sigma = f(\{Z_i\}),
\end{equation}
where $f(\dots)$ represents a generic functional form relating the angular dependence of the surface charge distribution and the positions of internal charges. This determines the charge distribution of the CSp model in such a way that it generates a surface potential identical to the one of its ICi counterpart (see SI for detailed derivations). Potential-matching the two models at the level of single particles results in two pair energy landscapes: {\em (i)} the interaction energy $V^\mathrm{ICi}$ between impermeable spheres with a discrete distribution of point charges in their interior (ICi) and {\em (ii)} the interaction energy $V^\mathrm{CSp}$ between permeable spheres (CSp) with a continuous surface charged distribution that has been tailored on the basis of the ICi counterpart. The pair interaction energies of the two models can now be compared, $V^\mathrm{ICi}\longleftrightarrow V^\mathrm{CSp}$, to verify whether and to what extent they match (Fig.~\ref{fig:approach}d).

\begin{figure*}
\centering  
\includegraphics[width=\textwidth]{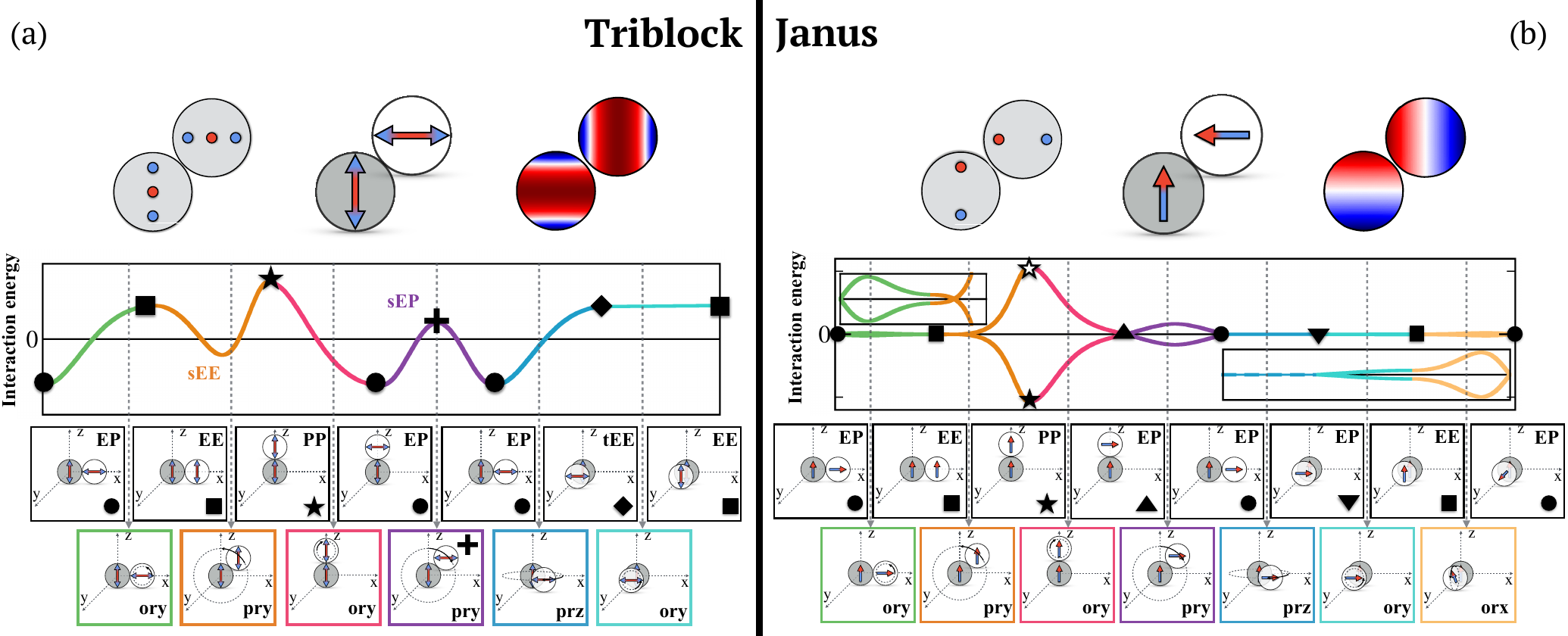}
\caption{Sketches of one-parametric orientational pathways tracing the interaction energy along the main orientations between two patchy particles at a fixed interparticle distance. {\bf (a)} Patchy particles with the symmetry of a linear quadrupole (Triblock) and {\bf (b)} patchy particles with the symmetry of a dipole (Janus, with two symmetric off-centre charges). The main orientations of the two particles relate the relative orientations of their equators (E) and poles (P): EP, EE, PP, tEE (transverse equator-equator), sEE (staggered equator-equator), and sEP (staggered equator-pole). Rotations (r) between configurations along the pathway either rotate one particle around the other (p) or the particle itself (o) along one of the three coordinate axes. In both panels, symbols label selected pair configurations and colors mark rotations between two consecutive ones along the chosen pathway. The empty star in the Janus pathway mark the PP configuration where one of the particles is flipped upside-down compared to the PP configuration marked by the full star. The interaction energy curves in panels (a) and (b) are sketches that qualitatively summarize the interparticle interactions.}
\label{fig:symmetries}
\end{figure*}

\subsection*{From potential-matching to pair interaction energy: homogeneously charged particles}

An instructive example of our framework is the case of homogeneously charged particles. In the ICi model, this corresponds to a single charge $Z$ placed at the centre of the particle; in the CSp model, this corresponds to a uniform surface charge distribution $\sigma(\Omega)=\sigma_0=Q/4\pi R^2$ imparting total charge $Q$ to the particle. Our approach to determine the effective charges used in the ICi pair potential in this case yields the DLVO excluded volume factor, as expected~\cite{siryk2021jcp,VerweyOverbeek-1948}. Matching the surface potentials of individual ICi and CSp particles leads to (see SI)
\begin{equation}
\label{eq:qz}
	\frac{Q}{Z}=\frac{\kappa R}{1+\kappa R} \frac{e^{\kappa R}}{\sinh(\kappa R)}.
\end{equation}
In other words, the total charge of the CSp particle $Q$ has to be modified in a $\kappa$-dependent fashion in order for the surface potentials to match. The resulting pair interaction energies then also match {\em exactly}, $V^\mathrm{IC}/V^\mathrm{CS}=1$. If one would instead equate the total charge on the ICi and CSp particles, $Z=Q$, this would lead to a $\kappa$-dependent difference in the interaction energies, $V^\mathrm{IC}/V^\mathrm{CS}=[\kappa R/(1+\kappa R)]^2\,[e^{2\kappa R}/\sinh^2(\kappa R)]$. 

This procedure can be applied to an arbitrary number and distribution of internal charges in the ICi model and a corresponding inhomogeneous surface charge distribution in the CSp model. While the potential-matching between the two models at the level of single particles is exact, the approximation of the pair energy involves calculation of effective charges $\{Z_i^*\}$ which only corrects for the excluded volume contribution up to a finite number of terms of the multipolar expansion (see SI). Evaluating the effects of this approximation for the effective charges is not straightforward as charge inhomogeneity also leads to intricate interaction landscapes of interparticle orientations that are not trivial to scan and compare. 

\subsection*{Orientational interaction landscapes of charged triblock and Janus patchy particles}

In colloids and proteins alike, the two types of charge anisotropy with the lowest symmetry -- that of a dipolar Janus particle and of a quadrupolar triblock particle -- still exhibit a rich and complex array of assembly and aggregation phenomena also seen in particles with more complex charge distributions and as such represent excellent model systems~\cite{hong2006clusters,naderi2020self}. Moreover, Janus and triblock charge distributions provide a good first-order approximation for charge distributions of globular proteins~\cite{hoppe2013simplified,Blanco_2016,bozic2017ph}. The pair interaction energy between these particles encapsulates the properties of the interactions in the system needed for both simulations and theoretical evaluations and in general depends on orientations and relative positions of both particles. Even for axisymmetric particles, the configuration space is three-dimensional and thus unfeasible to visualize in its entirety. In order to carry out a comparison between the pair interaction energies of CSp and ICi models, we construct a one-parametric rotational path that includes all highly-symmetric relative orientations of the two particles at a fixed interparticle distance $\rho$ and interpolates them in a continuous fashion. This method is inspired by Brillouin zone paths used to sample electronic band structures in crystal lattices. Figure~\ref{fig:symmetries} shows the rotational paths with corresponding sketches at most symmetric orientations, used for triblock particles with quadrupolar symmetry (panel (a)) and for Janus particles with dipolar symmetry (panel (b)). In the Janus case, we trace the path twice, with one of the particles reversed the second time around. Negative energies correspond to attraction in the radial direction and vice-versa, with expected maxima where like charges are in close proximity. The radial dependence of the interaction energy is less relevant because it will generally display an exponential fall-off as the interparticle distance increases. We thus typically study the orientational interaction landscapes when the two particles are in contact, $\rho=2R$. We also note that orientational paths such as the ones presented in Fig.~\ref{fig:symmetries} can be constructed for any charge distribution by choosing an appropriate set of mutual orientational configurations of particles and the transitions between them. This set does not need to be exhaustive (and in fact cannot be for complicated charge distributions) in order to characterize the orientational pair interaction profiles to a satisfying degree.

\subsection*{From potential-matching to pair interaction energy: triblock and Janus particles}

\begin{figure}[!ht]
\centering  
\includegraphics[width=\linewidth]{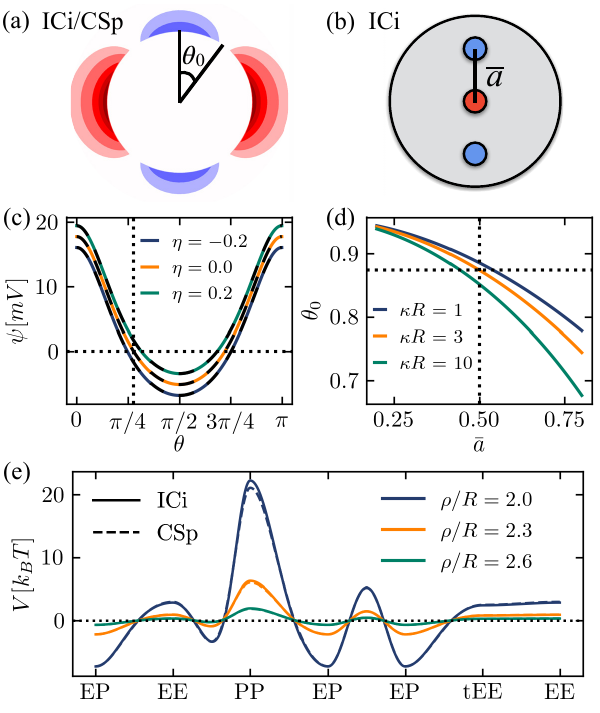}
\caption{Triblock symmetry. {\bf (a)} Sketch of the electrostatic potential around a triblock particle, with patch size $\theta_0$ defined as the angle of vanishing surface potential. {\bf (b)} Point charge distribution of a triblock particle in the ICi model, where $\bar a=a/R$ describes the relative displacement of positive (patch) charges $Z_p$ from the centre, referred to as $eccentricity$. {\bf (c)} Electrostatic potential of a single triblock particle with $R=150$ nm after potential-matching as a function of the polar angle, at $\kappa R=3$ and $\bar a=0.5$. The two models are depicted by solid (ICi) and dashed (CSp) lines. The potential is shown for particles with different total charge, given by parameter $\eta=Z_{tot}/|Z_p|$, where we used $Z_p = 280e$. Dotted lines highlight the patch size $\theta_0$ for neutral particles ($\eta=0$). {\bf (d)} Patch size $\theta_0$ as a function of the distance between the centre and off-centre charges, $\bar a$, in the ICi model at different values of $\kappa R$. {\bf (e)} Orientational pair interaction energy pathways for the ICi and CSp models, shown at three different interparticle distances $\rho$. The particles are overall neutral ($\eta=0$) and the other parameters are fixed ($\bar a=0.5$, $\kappa R=3$).}
\label{fig:mapping-triblok}
\end{figure}

We start by examining the interaction energies of triblock patchy particles, as their orientational pathway is simpler because of their symmetry (Fig.~\ref{fig:symmetries}a). To define the size of the charge patches on the particle poles in a consistent fashion, we use the polar angle $\theta_0$ at which the surface potential vanishes (Fig.~\ref{fig:mapping-triblok}a), which is the same for both models by construction (continuous versus dashed lines in Fig.~\ref{fig:mapping-triblok}c). The patch size $\theta_0$ in the ICi model is varied by changing the eccentricity $\bar a=a/R$ of the off-centre charges $Z_p$ (Fig.~\ref{fig:mapping-triblok}b). The patch size is the largest when the off-centre charges are close to the central one, and decreases in size as they are moved closer to the surface. This, in turn, increases the values of the potential at the poles. Two other important parameters influencing $\theta_0$ are the total charge on the particles, expressed by the dimensionless ratio $\eta=Z_\mathrm{tot}/|Z_p|$ (Fig.~\ref{fig:mapping-triblok}c) and the screening in the system given by $\kappa R$ (Fig.~\ref{fig:mapping-triblok}d). Going from zero net charge on the particle to negative or positive net charge acts to either decrease or increase $\theta_0$, respectively (Fig.~\ref{fig:mapping-triblok}c). As the total charge increases, it is possible to reach a limit beyond which the surface potential does not change sign anymore; in that case we do not consider the particles to have charged patches even if their surface potential is not isotropic. When the net particle charge is fixed, $\theta_0$ decreases upon increasing $\kappa R$ (Fig.~\ref{fig:mapping-triblok}d), i.e., patches become smaller when the screening in the system is higher.

\begin{figure*}[!ht]
\centering  
\includegraphics[width=\linewidth]{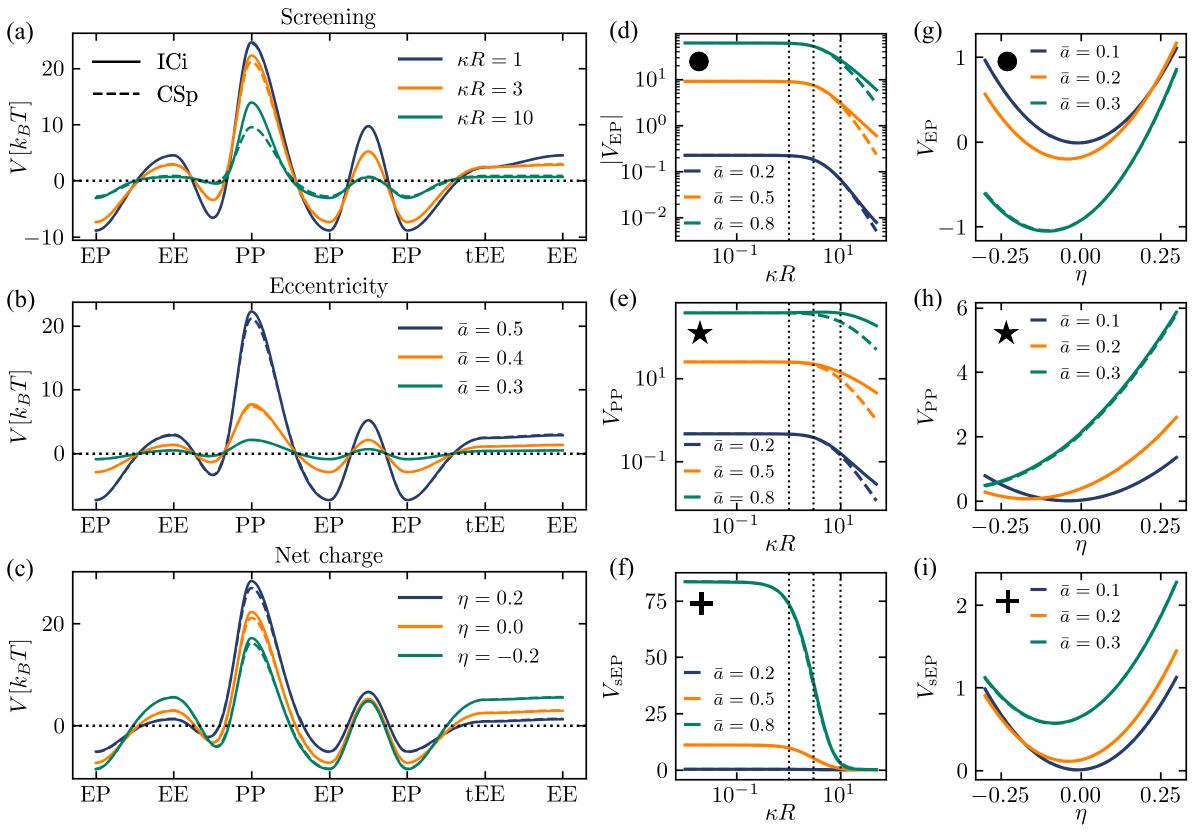}
\caption{Pair interaction energy of triblock particles in the ICi (full lines) and CSp (dashed lines) models. The particles are in contact, $\rho/R=2$, and we vary {\bf (a)} screening ($\kappa R$), {\bf (b)} patch size ($\bar{a}$), and {\bf (c)} net particle charge ($\eta$). Parameter values, when not varied, are $\kappa R = 3$, $\bar a=0.5$, and $\eta=0$. {\bf (d)}--{\bf (i)} Comparison of interaction energies in the two models at specific particle orientations between two particles (EP, PP, and sEP; cf.\ Fig.~\ref{fig:symmetries}) with (d)--(f) changing screening and (g)--(i) changing net particle charge. Dotted vertical lines in panels (d)--(f) highlight the values of the corresponding parameters in panels (a) and (c).}
\label{fig:potential-parameters-triblock-contact}
\end{figure*}

After mapping the single-particle electrostatic potential at the surfaces of triblock ICi and CSp particles, we observe that the resulting {\em pair} interaction energies show striking near-equivalence, as they are almost identical along the entire orientational pathway (Fig.~\ref{fig:mapping-triblok}e). This is true not only at contact but also at larger interparticle distances. Indeed, for most particle-particle orientations, the interaction energies in the two models are in excellent agreement (Fig.~\ref{fig:potential-parameters-triblock-contact}) regardless whether we vary the strength of screening  $\kappa R$ (panel (a)), the patch size by varying $\bar{a}$ (panel (b)), or the total charge on the particles $\eta$ (panel (c)). Differences between the two models only become apparent when the screening becomes large, in particular at the local extrema of the orientational pathway. We compare the absolute energy values of three extrema -- one minimum (EP) and two maxima (PP and sEP) -- over a large range of $\kappa R$ at fixed values of $\bar{a}$ for neutral particles (panels (d)-(f) of Fig.~\ref{fig:potential-parameters-triblock-contact}) and over a range of total charge $\eta$ at fixed values of $\bar{a}$ and $\kappa R$ (panels (g)-(i)). Differences between ICi and CSp models become visible only when $\kappa R\gtrsim 10$ as the effects of local charge patterns become more pronounced. This highlights the importance of higher-order terms in the potential expansion (Fig.~S2 in SI) as the determination of effective charges in the ICi model less accurately reproduces the features of the original potential, since only a finite number of expansion terms can be matched. This effect is the most pronounced at large $\bar a$ where patches are small in size and large in magnitude, especially in the PP orientation of the particles (Fig.~\ref{fig:potential-parameters-triblock-contact}e).

\begin{figure}[!ht]
\centering  
\includegraphics[width=\linewidth]{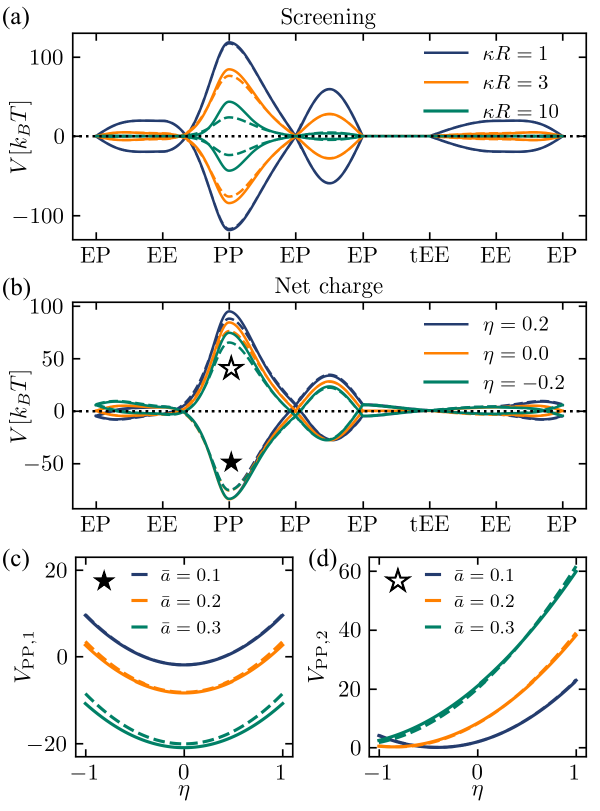}
\caption{Pair interaction energy of Janus particles in the ICi (full lines) and CSp (dashed lines) models. The particles are in contact, $\rho/R=2$, and we vary {\bf (a)} screening ($\kappa R$) and {\bf (b)} net particle charge ($\eta$). Parameter values, when not varied, are $\kappa R = 3$, $\bar a=0.5$, and $\eta=0$. Panels {\bf (c)} and {\bf (d)} show the dependence of PP peak energies on the total charge $\eta$ for both path branches (full and empty symbols, respectively).}
\label{fig:potential-parameters-janus-contact}
\end{figure}

Next, we consider the case of two Janus particles with off-centre charges in a similar fashion as we did for triblock particles. Due to the additional asymmetry of Janus particles compared to the triblock ones, the orientational pathway is more complicated and splits into two branches upon flipping one of the particles (full and empty symbols in Fig.~\ref{fig:symmetries}b). Despite this, once the Janus particles are potential-matched between the ICi and CSp models, their interaction energies exhibit the same degree of equivalence as they did in the triblock case (Fig.~\ref{fig:potential-parameters-janus-contact}). The discrepancies between the two models again become apparent only when the screening is large ($\kappa R\gtrsim 10$), and are most prominent in the interaction peaks and in the PP orientation of the two particles in particular (Fig.~\ref{fig:potential-parameters-janus-contact}a). Furthermore, as soon as the particles are not net neutral, the symmetry around zero between the two orientational pathway branches is lost due to the interaction of the additional monopole contribution with the (odd) higher order multipole expansion terms that change sign on particle reversal (Fig.~\ref{fig:potential-parameters-janus-contact}b). This effect is most obvious in the PP orientation where both branches show markedly different net charge dependence. 
In the lowest order, the monopolar contribution cancels out for parallel dipoles (full star, Fig.~\ref{fig:potential-parameters-janus-contact}c), while for antiparallel dipoles (empty star, Fig.~\ref{fig:potential-parameters-janus-contact}d) it results in additional attraction or repulsion of the dipoles, depending on its sign. For larger net charges, nontrivial higher order effects can be observed.

\subsection*{Discussion}

Our work demonstrates the connection between different mean-field representations of {\em pairwise} electrostatic interactions between charged patchy particles with their correspondence based on matching solely the {\em single-particle} surface electrostatic potentials. Using this approach, we are able to match IC and CS models to have the same interparticle interaction up to the accuracy that the two models can provide. The proposed framework serves as a self-consistent tool for designing and refining anisotropic DLVO-like models, suitable for large-scale simulations.

Both IC and CS models resort to similar simplifications. Foremost, they are derived in the linear DH approximation, with all its caveats, and they can be extended to the non-linear (PB) regime using charge renormalization~\cite{trizac2002simple,trizac2003alexander}. The two models further assume that the dielectric constant both inside and outside the particles is that of water, which is a simplification as systems of colloids or proteins are typically characterised by a dielectric jump. Our potential-mapping approach can be extended to dielectric particles by resorting to the known single-particle solution for the electrostatic potential of a dielectric CSp model~\cite{bozic2018jcp} and using it in the CS pair interaction energy to obtain a first-order approximation for the interaction of impermeable dielectric particles. Both models also assume perfectly spherical particles, which is not the case when modelling biological molecules such as proteins. This could be partially alleviated by using multiple spherical particles as a coarse-grained model of a more complex building block~\cite{pusara2021coarse}. To describe proteins, the assumption of fixed charge distributions that do not depend on pH should also be revisited. However, the inclusion of charge regulation or changes in particle conformation, both relevant for protein electrostatics~\cite{lund2005charge,zhou2018electrostatic}, presents a significant challenge, as it requires dynamic adjustments to the model in response to external stimuli. Nonetheless, once this challenge is addressed by developing a suitable model, our framework provides a method to assess the reliability of the model and test it against explicit electrostatic solutions.

The classic Janus and triblock charge distributions examined in this work can be realized with reasonable accuracy in synthetic charged patchy colloids by using state-of-the-art techniques. More importantly, they represent the leading terms of the multipole expansion that describes charge distributions of many globular proteins~\cite{hoppe2013simplified,Blanco_2016,bozic2017ph}. If more internal charges are needed to model a complex charge pattern, the IC and CS models can still be matched: when the internal charge positions and charges are already given, the corresponding surface charge distribution can be determined to exactly match the surface potential between the two models. More care is required in the calculation of effective charges that compensate for absent screening in the excluded volume within the ICi model. However, if a suitable finite subset of expansion terms can be found that best describe the potential features, the relation connecting the real and effective charges is linear and thus trivial to evaluate (Sec.~II of SI). A similar procedure can also be constructed for the inverse connection between the models, i.e., when the charge position can be inferred from the symmetries of the charge pattern, their magnitude could be determined from the CS multipole expansion coefficients.

The general inverse problem of finding an IC model for a known surface potential in a CS model -- relevant for modelling complex proteins -- is fully nonlinear and thus more challenging, because both optimal positions of internal charges as well as their number have to be determined. This requires an iterative procedure that is not limited to surface charge distributions, but could match volumetric charges without bringing much additional complexity. Obtaining a carefully selected point charge distribution as an approximation to a general charge distribution, permeable or impermeable, can thus be considered as its own step of the procedure; one such approach has been described by \citet{hoppe2013simplified}. Our results demonstrate that the single-particle (IC or CS) model obtained from this generalized problem will result in an accurate pair interaction energy and subsequently a well-behaved model of a large-scale system.

A particularly useful application of our framework is the design of coarse-grained models of proteins, which are typically characterized by intricate charge distributions. The introduced orientational pathway (Fig.~\ref{fig:symmetries}) provides a reference for coarse-grained models and allows to tune coarse-grained parameters to minimize the differences with respect to more detailed (and computationally expensive) models. Such a procedure would preserve the most important electrostatic features of modelled particles and enable simulation of large multi-particle systems.

\begin{acknowledgements}
EB and EL acknowledge support from the Austrian Science Fund (FWF) under Proj.\ No.\ Y-1163-N27. EL acknowledges support from the MIUR grant Rita Levi Montalcini. AB, SČ and AG acknowledge support from Slovenian Research Agency (ARIS) under contracts no.\ P1-0055 (AB), P1-0099 (SČ and AG), J1-50006 (SČ and AG) and J1-3027 (SČ). All authors thank Jeffrey Everts for his insights and suggestions.
\end{acknowledgements}

\section*{Methods}

\subsection*{Governing equations of ICi and CSp models}

In both ICi and CSp models, the particles are immersed in a dielectric solvent with dielectric constant $\varepsilon$ (we use that of water, $\varepsilon_w = 80$) in the presence of a symmetric monovalent salt with bulk concentration $c_0$. For simplicity, the dielectric constant inside the particle is assumed to be the same as outside and there is therefore no dielectric jump across the particle boundary.

We describe the system in terms of mean-field electrostatics and use linear Debye-H{\"u}ckel (DH) theory to derive the electrostatic potential $\psi(\mathbf{r})$ of two interacting particles:
\begin{equation}
\label{eq:dh}
    \nabla^2\psi=\kappa^2\psi,
\end{equation}
where, for sake of clarity and generality, we have dropped the dependency of the potential on spatial coordinates $\mathbf{r}$. Here, $\kappa=\sqrt{8\pi l_B c_0}$ is the inverse DH screening length which arises due to the screening of monovalent salt ions; $l_B=\beta e^2/4\pi\varepsilon\varepsilon_0$ is the Bjerrum length, with $\beta=1/k_BT$ and $e$ the elementary charge. 

The DH equation~\eqref{eq:dh} determines the electrostatic potential outside the two particles $\psi_\mathrm{out}^{\mathrm{IC}/\mathrm{CS}}$ in both ICi and CSp models. Furthermore, the same equation holds also for the electrostatic potential in the interior of the {\em permeable} particles of the CSp model, $\psi_\mathrm{in}^{\mathrm{CS}}$. On the other hand, the potential inside the {\em impermeable} particles of the ICi model must satisfy
\begin{equation}
\label{eq:ps}
    \nabla^2\psi^\mathrm{ICi}_\mathrm{in}=-\frac{\rho}{\varepsilon\varepsilon_0},
\end{equation}
where the charge density inside the particles $\rho$ is given in terms of the positions of the internal (point) charges $\{Z_i\}$.

\subsection*{Pair interaction energies}

The solutions for the pair interaction energies in both ICi and CSp models are known and have been derived previously~\cite{bianchi2011sm,bianchi:2015,bozic2013jcp}; here, we provide a brief summary.

\subsubsection*{Interaction energy in the ICi model}

We approximate the pair interaction energy of two ICi particles, $V^\mathrm{ICi}$, as a symmetrized energy of a probe particle in the field generated by a source particle~\cite{bianchi2011sm}. To properly carry out this calculation, the probe particle needs to contain effective charges $\{Z_i^*\}$ that take into account the lack of ions inside it, i.e., the excluded volume around the bare charges $\{Z_i\}$. Our derivation of the effective charges, given in the next Section, improves on the existing DLVO correction used previously in this model~\cite{bianchi2011sm,bianchi:2015}.

The probe particle with effective charges is placed in the {\em single-particle} external potential of an unmodified ICi particle. Since the pair energy between two particles must be symmetric, we symmetrize the pair interaction energy by including both the energy of the first particle in the field generated by the second particle and vice versa:
\begin{equation}
  \colorlet{oldcolor}{.}
  \color{black}
	V^\mathrm{ICi}=\frac{1}{2}\left[\sum_{i\in 2}\psi_1^\mathrm{ICi}({\bf r}_{i})\, Z_{i}^*+\sum_{i\in 1}\psi_2^\mathrm{ICi}({\bf r}_{i})\, Z_{i}^*\right].
\label{eq:int-ic}
\end{equation}
The form of the effective potential energy in Eq.~\eqref{eq:int-ic} is of the utmost importance from the viewpoint of simulations, as it can be used, for instance, as an effective Hamiltonian in Monte Carlo simulations.

\subsubsection*{Interaction energy in the CSp model}

To determine the pair energy of two CSp particles, we can derive the electrostatic potential of the system of two particles as a whole, $\Psi_{12}$~\cite{bozic2013jcp}. Due to the linearity of the DH equation, the electrostatic potential of the system is the sum of the individual potentials generated by each particle, namely:
\begin{equation}
    \Psi_{12}=A_1\psi_1^\mathrm{CSp}+A_2\psi_2^\mathrm{CSp},
\end{equation}
where each single-particle potential $\psi_i^\mathrm{CSp}$ is written in the coordinate system of its source particle $i$ and $A_i$ are (some) constants. The solution of the governing DH equation that determines the coefficients $A_i$ can be obtained by writing the electrostatic potential in the coordinate system of a single particle while satisfying the boundary conditions on both particles simultaneously. To achieve this, one can resort to addition theorems for the potentials $\psi_i^\mathrm{CSp}$, resulting in an {\em exact} analytical expression for the electrostatic potential of the entire system of two particles in the coordinate system of either the first or the second particle, namely, $\widetilde{\Psi}_1$ or $\widetilde{\Psi}_2$.

The free energy of this system is obtained via the charging integral over the surface of both charged particles:
\begin{equation}
    F^\mathrm{CSp}=\frac{1}{2}\sum_{i=1,2}\oint_{S_i}\sigma_i\widetilde{\Psi}_i\Big|_{r=R_i}\mathrm{d}S_i,
\end{equation}
where the integral runs over the closed surface of each particle. Finally, the interaction energy between two CSp particles is obtained as the free energy required to bring them from infinite separation to a distance $\rho$:
\begin{equation}
V^\mathrm{CSp}=F^\mathrm{CSp}(\rho)-F^\mathrm{CSp}(\infty).
\end{equation}
This leads to a free energy of pairwise particle interaction that can be written as a multipole expansion~\cite{bozic2013jcp}.

\subsection*{Effective charges of probe particle in the ICi model}

We determine the effective charges $Z_i^*$ of the probe particle within the IC model by requiring that the single-particle electrostatic potential at the particle surface of an impermeable sphere is equal to the potential of its permeable counterpart, namely
\begin{equation}
  \colorlet{oldcolor}{.}
  \color{black}
	\psi^\mathrm{ICp}\Big|_{r=R} =  \psi^\mathrm{ICi}\Big|_{r=R} 
\Rightarrow
  \colorlet{oldcolor}{.}
  \color{black}
  \color{oldcolor}{ \forall i \hspace{0.1em} Z_i^* =g(\{Z_i\})}.
  \end{equation}
As both potentials can be expressed through multipole expansions, the aforementioned mapping consists of matching the expansion coefficients. While the equivalence is thus exact for an infinite number of terms, we consider only the leading ones, a choice that depends on the number and distribution of the point charges. Detailed derivations for charge distributions with different symmetries are given in Sec.~II of SI.

\clearpage
\newpage

\onecolumngrid

\setcounter{equation}{0}
\setcounter{figure}{0}
\setcounter{table}{0}
\setcounter{page}{1}
\setcounter{section}{0}
\setcounter{page}{1}
\makeatletter
\renewcommand{\theequation}{S\arabic{equation}}
\renewcommand{\thefigure}{S\arabic{figure}}
\renewcommand{\thetable}{S\arabic{table}}
\renewcommand{\thesection}{S\arabic{section}}
\renewcommand{\thepage}{S\arabic{page}}

\onecolumngrid

\widetext
\begin{center}
\textbf{\Large  \\ \vspace*{1.5mm} SUPPORTING INFORMATION: \\ Anisotropic DLVO-like interaction for charge patchiness in colloids and proteins} \\
\vspace*{5mm}
Andra{\v z} Gnidovec, Emanuele Locatelli, Simon {\v C}opar, An{\v z}e Bo{\v z}i{\v c}, Emanuela Bianchi
\vspace*{10mm}
\end{center}

\section{Matching the surface potentials of single particles in both models.}
\label{sec:surface-charge}

\begin{figure}[!ht]
\centering  
\includegraphics[trim={0.25cm 1cm 0.25cm 0.5cm},clip,width=0.7\textwidth]{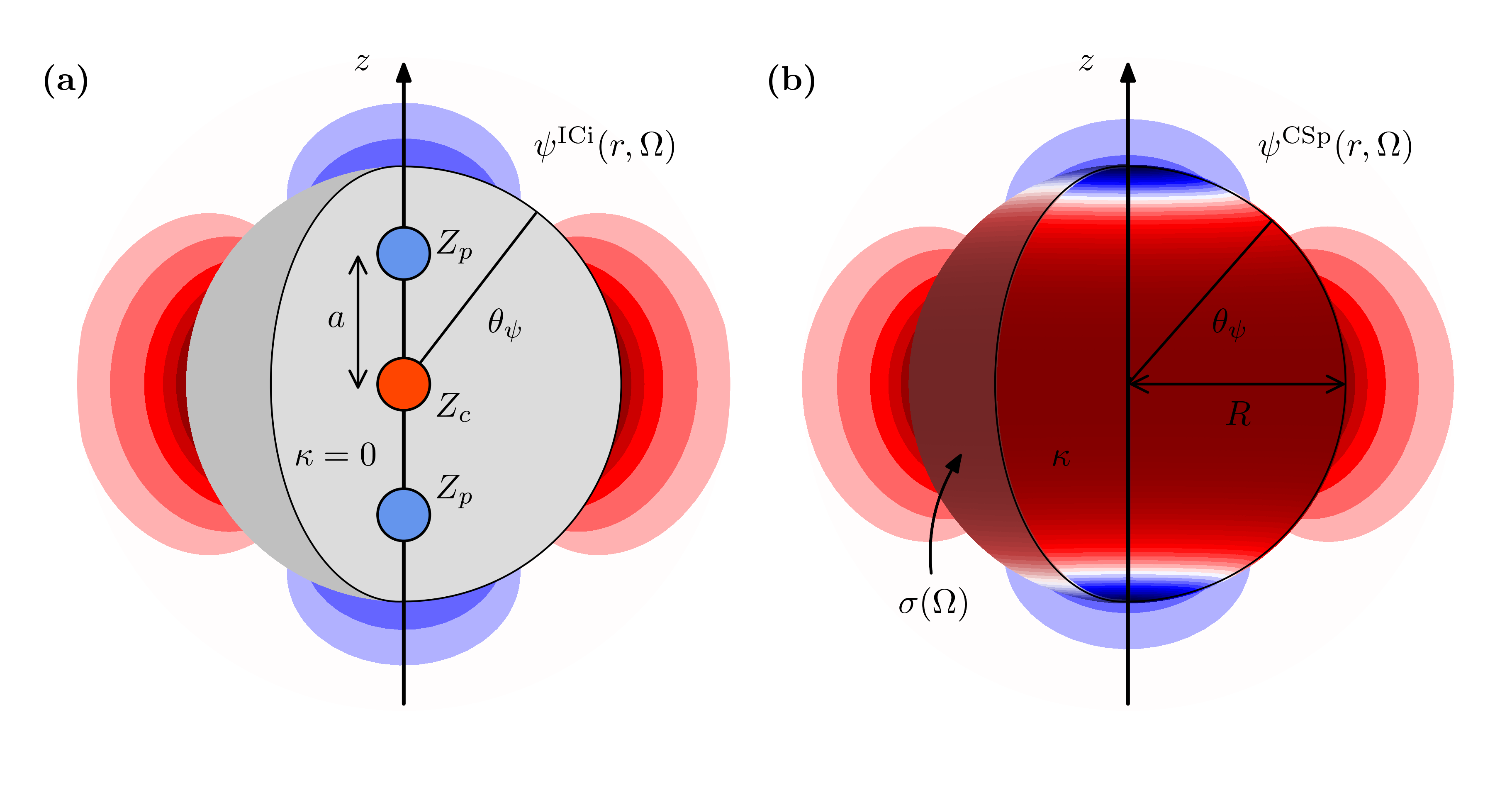}
\caption{Sketch of the ICi and CSp models applied to a patchy particle with the symmetry of a linear quadrupole (triblock particle). {\bf (a)} Internal charge (ICi) model and {\bf (b)} charged shell (CSp) model. In both models, the exterior of the particles consists of a symmetric monovalent salt with bulk concentration $c_0$, giving rise to the inverse screening length $\kappa$. The ICi model in panel (a) consists of a central charge $Z_c$ and two eccentric charges $Z_p$ positioned at a distance $\pm a$ along the $z$-axis inside of a sphere with radius $R>a$. The sphere is not permeable to salt ions, $\kappa=0$. The CSp model in panel (b) consists of a permeable sphere of radius $R$ carrying a surface charge density $\sigma(\Omega)$. Each of the two panels shows the electrostatic potential outside of the particle, $\psi_\mathrm{out}$. The ``patches'' of charge at the two poles of the particles are defined through the zero of the surface potential (patch angle $\theta_\psi$).}
\label{fig:models-sketch-triblock}
\end{figure}

Both types of particles considered in our work -- triblock and Janus particles -- have an axisymmetric charge distribution. While the charge distribution of Janus particles has the symmetry of a dipole, the charge distribution on triblock particles has a symmetry of a linear quadrupole (Fig.~\ref{fig:models-sketch-triblock}). In the ICi model, both examples are realized with three colinear charges inside the particle, with one charge of magnitude $Z_c$ lying in the particle center and the other two \emph{patch} charges, $Z_{p1}$ and $Z_{p2}$, displaced by $\pm a$ from the central charge. We impose equal magnitude of patch charges $|Z_{p1}|=|Z_{p2}|$, where in the Janus case they are of opposite signs, $Z_{p1}=-Z_{p2}=Z_p$, and of equal signs $Z_{p1}=Z_{p2}=Z_p$ in the Triblock case. The particle is overall neutral if the sum of all the charges is zero and carries a total charge $Z_\text{tot}=Z_c+Z_{p1}+Z_{p2}$ otherwise. The ICi particle itself is impermeable to salt ($\kappa=0$). In the CSp model, both Janus and Triblock particles are modelled as having a surface charge distribution $\sigma(\Omega)$ such that it results in the appropriate symmetry, meaning that the dominant term in its multipole expansion will either be of the order $l=1$ for Janus particles or of the order $l=2$ for Triblock particles (Fig.~\ref{fig:models-sketch-triblock}b). A non-zero net charge of the particle can be modelled by simply adding a term with $l=0$. The CSp particles are permeable to salt ions.

The electrostatic potential $\psi^{\rm ICi}$ outside the ICi particle in the DH approximation for both the Janus and the Triblock case can be written as~\cite{bianchi2011sm}:
\begin{equation}
\psi^\mathrm{ICi} (r, \theta) =  \psi_0(r) + \frac{1}{\kappa\varepsilon\varepsilon_0}\frac{2Z_p}{4\pi R^2} 
\sum_{l>0}\phantom{}' (2l+1)\bar{a}^l \frac{k_{l}(\kappa r)}{k_{l+1}(\kappa R)} P_l(\cos\theta),
\label{eq:psiIC}
\end{equation}
where $k_l(x)$ are the modified spherical Bessel functions of the second kind and $\sum'$ should be taken over all odd $l$ in the Janus case and over all even $l$ for Triblock particles. Additionally, $\psi_0$ is the spherically symmetric ($l=0$) term
\begin{equation}
    \psi_0(r)=\frac{Z_\text{tot}}{4\pi\varepsilon\varepsilon_0}\frac{e^{\kappa R}}{1+\kappa R}\frac{e^{-\kappa r}}{r},
\end{equation}
which equals to zero for a neutral particle.
In the CSp model, the external solution is again given in terms of $k_l(x)$~\cite{bozic2013jcp},
\begin{equation}
    \psi^\mathrm{CSp}(r,\Omega)=\frac{1}{\kappa\varepsilon\varepsilon_0}\sum_{l,m}C_\mathrm{p}(l,\kappa R)\frac{k_l(\kappa r)}{k_l(\kappa R)}\sigma_{lm}Y_{lm}(\Omega),
    \label{eq:psiCS}
\end{equation}
where
\begin{equation}
\label{eq:cp}
    C_\mathrm{p}(l,x)=x^2i_l(x)k_l(x),
\end{equation}
and $\sigma_{lm}$ are the multipole expansion coefficients of the surface charge density $\sigma(\Omega)$. Note that one can easily write the solution of the CSp model when the sphere is impermeable to ions by replacing the function $C_\mathrm{p}(l,x)$ with~\cite{bozic2018jcp}
\begin{equation}
\label{eq:ci}
    C_\mathrm{i}(l,x)=\frac{k_l(x)}{k_{l+1}(x)},
\end{equation}
which becomes important when comparing the IC and CSp models. Matching the two potentials of the ICi and CSp models, Eqs.~\eqref{eq:psiIC} and~\eqref{eq:psiCS}, we obtain
\begin{equation}
\label{eq:sigma-corr}
    \sigma_{l0}=\sqrt{4\pi(2l+1)}\frac{C_\mathrm{i}(l,\kappa R)}{C_\mathrm{p}(l,\kappa R)}\frac{2Z_p}{4\pi R^2}\,\bar{a}^l.
\end{equation}
By choosing the surface charge distribution of the CSp model through the expansion coefficients given in Eq.~\eqref{eq:sigma-corr}, the outside potentials of the ICi and CSp models match exactly -- {\em everywhere} -- despite some fundamental differences between them.

Our approach provides a natural way to extend our potential matching also to dielectric particles: since the single-particle solution for the electrostatic potential of a dielectric CSp model is known exactly~\cite{bozic2018jcp}, one needs to replace the ratio of the functions $C_\mathrm{i}/C_\mathrm{p}$ with a different ratio $C_\mathrm{d}/C_\mathrm{p}$, where
\begin{equation}
\label{eq:cd}
C_\mathrm{d}(l,\kappa R)=\frac{1}{\kappa R}\left[\frac{l}{\kappa R}\left(\frac{\varepsilon_p}{\varepsilon_w}-1\right)-\frac{k_{l+1}(\kappa R)}{k_l(\kappa R)}\right]^{-1}
\end{equation}
characterises the single-particle electrostatic potential of a dielectric particle with dielectric constant $\varepsilon_p$. The single-particle solution can then be used in the pair interaction energy of the CS model to obtain a first-order approximation for the interaction of impermeable dielectric particles.


\section{Effective charges of probe particle in the IC\lowercase{i} model}
\label{sec:effchr}

The interaction energy of two patchy charged particles in the ICi model is determined as the symmetrized interaction of a {\em permeable} probe particle with effective charges (taking into account the excluded volume correction) in the external potential from the original {\em impermeable} particle (see Fig.~\ref{fig:approach} in the main text). We determine the effective point charges of the permeable probe particle (ICp) in such a what that its electrostatic surface potential coincides with the electrostatic surface potential of its impermeable counterpart (ICi). To achieve this, we express both ICp and ICi potentials in terms of Legendre polynomials and subsequently equate them term by term up to as many multipole terms as there are charges -- namely, up to $l=0$ (for homogeneously charged particles), $l=1$ (for Janus particles), and $l=2$ (for triblock particles).

\paragraph*{Homogeneously charged particles.} First, we consider the case of a single internal charge $Z$, possibly displaced by distance $a$ with respect to the center of the particle of radius $R$. The single-particle electrostatic potential at the surface of an impermeable particle with a single internal charge $Z$ is~\cite{hoffmann2004molphys}
\begin{equation}
     \psi^{\text{ICi}}(r,\theta)\Big|_{r=R}=\frac{1}{\kappa \varepsilon\varepsilon_0}\frac{Z}{4\pi R^2}\sum_l (2l+1)\frac{k_l(\kappa R)}{k_{l+1}(\kappa R)}\bar{a}^l P_l(\cos\theta),
     \label{eq:psiICI}
\end{equation}
where we introduced $\bar{a}=a/R$ and $k_l$ are the modified spherical Bessel functions of the second kind as defined by Abramowitz and Stegun~\cite{abramowitz1972}. On the other hand, the single-particle electrostatic potential at the surface of a permeable particle with a single -- possibly off-center -- effective internal charge $Z^*$ is
\begin{equation}
    \psi^{\text{ICp}}(r,\theta)\Big|_{r=R}=\left.\frac{Z^\ast }{4\pi \varepsilon\varepsilon_0}\frac{e^{-\kappa |{\bf r}-{\bf a}|}}{|{\bf r}-{\bf a}|}\right|_{r=R}=\frac{Z^\ast }{4\pi\varepsilon\varepsilon_0}\frac{e^{-\kappa \sqrt{R^2-2aR \cos\theta +a^2}}}{\sqrt{R^2-2aR \cos\theta +a^2}},
    \label{eq:psiICP-a}
\end{equation}
where $i_l$ are the modified spherical Bessel functions of the first kind~\cite{abramowitz1972}. Expanding Eq.~\eqref{eq:psiICP-a} in terms of Legendre polynomials, we get
\begin{eqnarray}
    \psi^{\text{ICp}}(r,\theta)\Big|_{r=R}
    &=&\sum_l \frac{2l+1}{2}\int_{-1}^{1} \psi^{\text{ICp}}(r,\theta)|_{r=R} P_l(\cos\theta){\rm d}(\cos\theta) \nonumber \\
    &=& \frac{1}{\kappa \varepsilon\varepsilon_0}\frac{Z^\ast}{4\pi R^2}\sum_l (2l+1)(\kappa R)^2 i_l(\kappa a)k_l(\kappa R) P_l (\cos\theta).
    \label{eq:psiICP}
\end{eqnarray}
We now compare the $l=0$ terms of the two potentials to obtain the equality
\begin{equation}
     Z\,\frac{k_0(\kappa R)}{k_1(\kappa R)} = Z^* (\kappa R)^2 i_0(\kappa a) k_0(\kappa R)
\end{equation}
which leads to
\begin{equation}
    Z^* = Z\frac{1}{k_1(\kappa R)i_0(\kappa a)(\kappa R)^2} = Z\frac{e^{\kappa R}}{\kappa R +1} \frac{\kappa a}{\sinh(\kappa a)}.
\end{equation}
When the charge is positioned in the center of the particle ($a=0$), this reduces to
\begin{equation}
    Z^* \underset{a\rightarrow 0}{\longrightarrow}Z\frac{e^{\kappa R}}{\kappa R +1},
    \label{eq:eff-dlvo}
\end{equation}
the DLVO limit corresponding to a single central point charge $Z$~\cite{VerweyOverbeek-1948}.

\paragraph*{Janus particles.} To derive the effective charge for the case of Janus particles (Fig.~\ref{fig:symmetries} in the main text), we consider three internal charges placed along the particle diameter. By choosing the two off-center charges to satisfy $Z_{p,1} = -Z_{p,2}$, we impose a dipole moment to the particle. We can further tune the overall (net) charge on the particle with a central contribution $Z_c$ (be it positive or negative).
Notice that given $\eta = Z_T/|Z_p|$, we have $Z_c = \eta |Z_p|$. By imposing equality between the zeroth and first moment of the expansions in Eqs.~(\ref{eq:psiICI}) and~(\ref{eq:psiICP}), we obtain the effective charges $Z^*_{p}$ and $Z^*_c$ :
\begin{eqnarray}
   Z^*_c &=& \frac{Z_c}{(\kappa R)^2} \frac{1}{k_1(\kappa R)} = Z_c\frac{e^{\kappa R}}{\kappa R +1} \quad\mathrm{and} \\
   Z^*_p &=& \frac{Z_p}{(\kappa R)^2} \frac{\bar{a}}{k_2(\kappa R) \, i_1(\kappa a)}.     
\end{eqnarray}

\paragraph*{Triblock particles.} We consider again three internal charges placed linearly along the particle diameter 
(Fig.~\ref{fig:symmetries} in the main text). Because of the cylindrical symmetry of the point charge distribution, we can still rely on Legendre polynomials rather than on spherical harmonics. As a consequence, we can simply substitute $Z$ and $Z^*$ with $\{Z_c,2Z_p\}$ and $\{Z^*_c,2Z^*_p\}$ respectively, and add the contributions together. Here $Z_c$ is the central charge and $Z_p$ are the two off-center charges with opposite sign. We thus obtain the following system of equations for $l=0$:
\begin{equation}
Z_c\frac{1}{k_1(\kappa R)} +2Z_p\frac{1}{k_1(\kappa R)} = Z^*_c(\kappa R)^2 + 2Z^*_p(\kappa R)^2i_0(\kappa a),
\end{equation}
and $l=2$:
\begin{equation}
Z_p\frac{\bar{a}^2}{k_3(\kappa R)} = Z^*_p(\kappa R)^2i_2(\kappa a),
\end{equation}
while the term $l=1$ is zero due to the symmetry of the problem. This leads to
\begin{eqnarray}
    Z_c^\ast &=&(Z_c+2Z_p)\frac{e^{\kappa R}}{1+\kappa R} - 2Z_p\frac{\bar{a}^2}{(\kappa R)^2k_3(\kappa R)}\frac{i_0(\kappa a)}{i_2(\kappa a)}\quad\mathrm{and}\\
    Z_p^\ast &=&Z_p\frac{\bar{a}^2}{(\kappa R)^2k_3(\kappa R) i_2(\kappa a)}.
\end{eqnarray}

\paragraph*{General charge distribution.} The above procedure for the calculation of effective charges can be generalized to arbitrary point charge distributions inside the impermeable particle. This calls for rewriting the potential expansion for a single displaced charge in terms of spherical harmonics instead of Legendre polynomials, in turn allowing us to independently rotate and combine single-charge expansions of all charges inside the particle. We thus get a general expansion form for the electrostatic potential of an impermeable particle with $N$ point charges,
\begin{equation}\label{eq:ICi-general}
     \psi^{\text{ICi}}\Big|_{r=R}=\sum_{l m} \left(\sum_i Z_i D_{0, m}^{(l)}(\hat{\bm{a}}_i) b_{i, l 0} \right) Y_{l m},
\end{equation}
where the sum over $i$ runs over all charges, $D_{0, m}^{(l)}(\hat{\bm{a}}_i)$ are Wigner D-matrices that rotate the original expansion in the direction of each individual charge $\hat{\bm{a}}_i$, and 
\begin{equation}\label{eq:ICi-coef}
    b_{i,l 0} = \frac{1}{R^2\kappa\varepsilon\varepsilon_0} \sqrt{\frac{(2l+1)}{4\pi}} \frac{k_l(\kappa R)}{k_{l+1}(\kappa R)} \bar{a}^l
\end{equation}
are original single-charge expansion coefficients from \eqref{eq:psiICI}, adjusted with the conversion factor $\sqrt{4\pi/(2l + 1)}$ from Legendre polynomials to spherical harmonics.

An equivalent expression to \eqref{eq:ICi-general} can be derived from Eq.~\eqref{eq:psiICP} for the potential of a permeable IC particle, $\psi^{\text{ICp}}$, with point charges $Z^*_i$ and original expansion coefficients
\begin{equation}\label{eq:ICp-coef}
    c_{i,l 0} = \frac{1}{R^2\kappa\varepsilon\varepsilon_0} \sqrt{\frac{(2l+1)}{4\pi}} (\kappa R)^2 i_l(\kappa a) k_l(\kappa R).
\end{equation}
Comparing $N$ expansion terms for $(l,m)$ pairs that most accurately describe the charge distribution, we get a linear system of equations that relate effective charge magnitudes to the original ones,
\begin{equation}
    Z^*_i = C^{-1}_{ij} B_{jk} Z_k,
\end{equation}
where $B_{ij} = D_{0, m}^{(l)}(\hat{\bm{a}}_i)\, b_{i, l 0}$ and $C_{ij} = D_{0, m}^{(l)}(\hat{\bm{a}}_i)\, c_{i, l 0}$, with index $j$ indicating $(l,m)$ pairs. The question of choosing an appropriate set of $(l,m)$ values is beyond the scope of this discussion; however, it should allow for the matrix $C$ to be invertible.

\begin{figure*}[!ht]
\centering  
\includegraphics[width=0.8\textwidth]{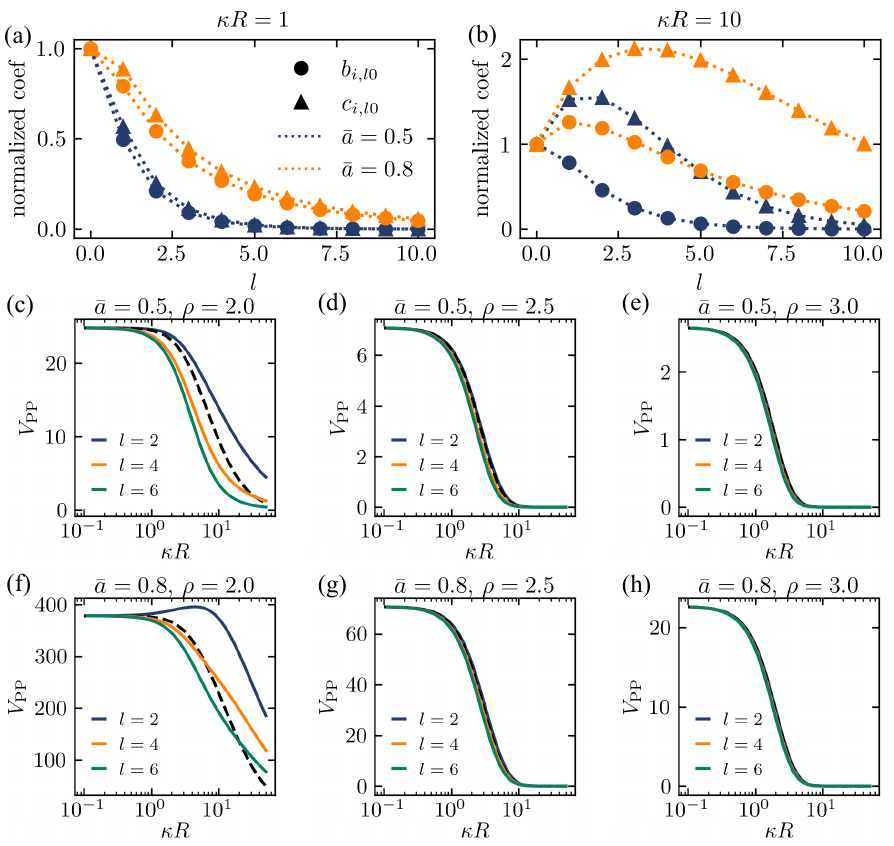}
\caption{Higher-order correction effects. (a-b) ICi and ICp expansion coefficients $b_{i,l0}$ and $c_{i,l0}$ for a single offset charge, Eq.~\eqref{eq:ICi-coef} and Eq.~\eqref{eq:ICp-coef}, normalized with their respective $l=0$ values. The panels show results at two different screening values, $\kappa R=1$ and $\kappa R = 10$. (c-h) Pair energy curves in the PP configuration of two triblock particles as a function of screening, at different distances $\rho$ and charge eccentricities $\bar a$. Dashed lines show CSp model results and full lines show ICi model results when different expansion terms are matched in the effective charge calculation.}
\label{fig:higher-term-matching}
\end{figure*}

\paragraph*{Alternative higher-order matching criteria.} The above derivation of effective charges for the Janus and triblock cases assumes that the most significant contributions to potential expansions are the dipole ($l=1$) and quadrupole ($l=2$) terms, respectively. This is not necessarily the case as expansion coefficients $b_{i,l0}$ [Eq.~\eqref{eq:ICi-coef}] and $c_{i,l0}$ [Eq.~\eqref{eq:ICp-coef}] depend significantly on both eccentricity $\bar a$ and screening $\kappa R$, as demonstrated in panels (a) and (b) of Fig.~\ref{fig:higher-term-matching}. In particular, for cases where off-center charges are close to the particle surface and at large screening, higher-$l$ terms can become larger than lower-$l$ terms. This effect is more pronounced for the ICp expansion, which leads to large deviations between the higher order terms of the impermeable and permeable cases. In order to achieve a closer match between ICi and ICp potentials, equality should be imposed between the higher order terms instead of the dipolar or quadrupolar ones at the cost of some deviation in lower order contributions, or an overdetermined system be solved in a least squares fashion to minimize a chosen cost function. In Fig.~\ref{fig:higher-term-matching}c-h, we demonstrate on the case of triblock particles that imposing equality between higher order terms in the calculation of effective charges can improve the match with the CSp model, especially at high $\bar a$ where matching the quadrupolar terms ($l=2$) leads to an unphysical increase in PP peak energy with the increase in screening strength (Fig.~\ref{fig:higher-term-matching}f). Nevertheless, equating higher order terms often only improves model matching at higher screening values while introducing larger deviations at smaller $\kappa R$, see e.g. Fig.~\ref{fig:higher-term-matching}c. Note also that energy differences between the ICi and CSp models at different matched expansion terms quickly vanish with increasing interparticle distance. 

The approach outlined in this section generalizes the process of assigning effective charges as known from the DLVO theory to an arbitrary number of point charges inside an impermeable particle by means of potential-matching at the level of multipole expansion. Further generalizations are possible, for instance, matching the potential between electrolytes with different screening lengths (different $\kappa$), of which the impermeable particle model represents the limit of $\kappa\to 0$.


\section{Interaction energy of two homogeneously charged particles}
\label{sec:monopoles}

The case of two homogeneously charged particles is used to test the robustness and self-consistency of our approach  (cf.\ Fig.~\ref{fig:approach} in the main text). To this purpose, we demonstrate here all the necessary steps to obtain Eq.~(2) in the main text.

\paragraph*{Interaction energy in the ICi model.} The electrostatic potential of a single isolated ICi particle with a central internal charge $Z$ is obtained from Eq.~\eqref{eq:psiICI} as $a \to 0$,
\begin{equation}
	\psi^\mathrm{ICi} (r) =  \frac{Z}{4\pi\varepsilon\varepsilon_0}\frac{e^{\kappa R}}{1+\kappa R}\frac{e^{-\kappa r}}{r},
\label{eq:psi-ICi-iso}
\end{equation}
To determine the pair interaction energy between two such particles at a distance $\rho$, the ICi model determines the energy of a probe particle -- {\em permeable} sphere of radius $R$ and (central) effective charge $Z^*$ [Eq.~\eqref{eq:eff-dlvo}] -- in the potential of the first particle [Eq.~\eqref{eq:psi-ICi-iso}],
\begin{equation}
    V^{\mathrm{ICi}}=Z^*\psi^\mathrm{ICi}(\rho)
\end{equation}
We thus obtain the pair interaction energy between two homogeneously charged ICi particles at a distance $\rho$,
\begin{equation}\label{eq:pair-ICi-Iso}
	V^\mathrm{ICi}(\rho)= \frac{Z^2}{4\pi\varepsilon\varepsilon_0}\left(\frac{e^{\kappa R}}{1+\kappa R}\right)^2\frac{e^{-\kappa \rho}}{\rho},
 \end{equation}
which is the well-known Yukawa potential as described by the DLVO theory~\cite{VerweyOverbeek-1948}.

\paragraph*{Interaction energy in the CSp model.} Electrostatic potential of a pair of homogeneously charged CSp particles at a point $\mathbf{r}$ outside both particles is written in a general form as~\cite{bozic2013jcp}
\begin{equation}
    \Psi(1,2)=A_1\frac{e^{-\kappa|\mathbf{r}-\mathbf{r}_1|}}{|\mathbf{r}-\mathbf{r}_1|}+A_2\frac{e^{-\kappa|\mathbf{r}-\mathbf{r}_2|}}{|\mathbf{r}-\mathbf{r}_2|}.
\end{equation}
Here, $A_1$ and $A_2$ are yet unknown coefficients which are determined by solving the DH equation [Eq.~\eqref{eq:dh} in the main text] with the appropriate boundary conditions of potential continuity and discontinuity of the electric field due to the charge on the particle surfaces. Using the addition theorem for spherical Bessel functions of the first and second kind, one is able to write out the pair electrostatic potential in the coordinate system of only one of the particles~\cite{bozic2013jcp}. Choosing the coordinate system of particle 2, we can write
\begin{equation}
    \widetilde{\Psi}(2)=\frac{\sigma_{0}}{\sqrt{4\pi}}\frac{\sinh\kappa R}{\kappa\varepsilon\varepsilon_0}\left[\frac{R}{r}\,e^{-\kappa r}+
    \frac{R}{r}\,\sinh\kappa r\,\frac{e^{-\kappa\rho}}{\kappa\rho}\right]
\end{equation}
(where one could equivalently opt for $\widetilde{\Psi}(1)$ in the coordinate system of particle 1). The free energy of the system of two particles at a distance $\rho$ is obtained via the charging integral over the surfaces of the two particles,
\begin{equation}
    F^\mathrm{CSp}=\frac{1}{2}\sum_{i=1,2}\oint_{S_i}\sigma_i \widetilde\Psi(i)\Big|_{r=R}\,\mathrm{d}S_i.
\end{equation}
The interaction energy of the two particles is then simply the free energy of particles at a distance $\rho$ brought over from infinity,
\begin{equation}
    V^\mathrm{CSp}=F^\mathrm{CSp}(\rho)-F^\mathrm{CSp}(\infty),
\end{equation}
where $F^\mathrm{CSp}(\infty)$ is simply the self-energies of both individual particles,
\begin{equation}
    \frac{1}{2}F^\mathrm{CSp}(\infty)=\frac{R^2}{2\kappa\varepsilon\varepsilon_0}\frac{4\pi\sigma_0^2}{1+\coth\kappa R}.
\end{equation}
For two identical homogeneously charged particles, we thus obtain their interaction energy as~\cite{bozic2013jcp}
\begin{equation}
\label{eq:homint}
    V^\mathrm{CSp}(\rho)=\frac{4\pi R^2 \sigma_0^2}{\kappa\varepsilon\varepsilon_0} \sinh^2(\kappa R)\frac{e^{-\kappa \rho}}{\kappa \rho}.
\end{equation}

\paragraph*{Matching the surface potentials of single particles in the ICi and CSp models.} The electrostatic potential outside an ICi particle is given by Eq.~\eqref{eq:psi-ICi-iso}, where $Z$ is the value of a point charge placed in the center of the impermeable particle. The electrostatic potential outside a single CSp particle is, on the other hand~\cite{bozic2013jcp}:
\begin{align}
	\psi^\mathrm{CSp}(r) &=  \frac{1}{\kappa\varepsilon\varepsilon_0}\sinh(\kappa R)e^{-\kappa R}\frac{k_0(\kappa r)}{k_0(\kappa R)}\sigma_0 \\ 
     &= \frac{R^2 \sigma_0}{\varepsilon\varepsilon_0} \frac{\sinh{(\kappa R)}}{\kappa R} \frac{e^{-\kappa r}}{r},
	\label{eq:psi-CSp-iso}
\end{align}

where $\sigma_0=Q/(4\pi R^2)$ is the constant surface charge of the CSp particle and $k_0(x)$ are the $0$-th order modified spherical Bessel functions of the second kind. Equating Eqs.~\eqref{eq:psi-ICi-iso} and~\eqref{eq:psi-CSp-iso} on the particle surface, we obtain the relationship between the surface charge of the CSp model and the point charge of the ICi model,
\begin{equation}\label{eq:sigma-to-z}
	\sigma_0=\frac{Z}{4\pi R^2} \frac{\kappa R}{1+\kappa R} \frac{e^{\kappa R}}{\sinh(\kappa R)}.
\end{equation}
In other words, $Q \neq Z$ and the factor
\begin{equation}\label{eq:qz_si}
	\frac{Q}{Z}=\frac{\kappa R}{1+\kappa R} \frac{e^{\kappa R}}{\sinh(\kappa R)}
\end{equation}
accounts for the difference between permeable and impermeable particles and makes the electrostatic potential on their surfaces identical, $\psi^\mathrm{ICi}|_{r=R}=\psi^\mathrm{CSp}|_{r=R}$.

We note that the charge ratio in Eq.~\eqref{eq:qz_si} can be succinctly expressed as a ratio of two functions,
\begin{equation}\label{eq:qzc}
    \frac{Q}{Z}=\frac{C_i(0,\kappa R)}{C_p(0,\kappa R)},
\end{equation}
where $C_i(l,\kappa R)$ and $C_p(l,\kappa R)$ characterize the $l$-th order multipole terms of the solutions for the single-particle electrostatic potential in the charged shell model when the shell is either impermeable or permeable to the salt ions, respectively (see Eqs.~\eqref{eq:cp} and~\eqref{eq:ci}).

\paragraph*{Comparison of the pair interaction energies in the two models.}

Using the potential matching to obtain the relationship between the charge on the CSp particles and the charge on the ICi particles [Eq.~\eqref{eq:sigma-to-z}], we can write the surface charge $\sigma_0$ as a function of $Z$, leading to the interaction energy of the CSp particles of the form [cf.\ Eq.~\eqref{eq:homint}]
\begin{equation}\label{eq:pair-CSp-Iso}
 	V^\mathrm{CSp}(\rho)= \frac{Z^2}{4\pi\varepsilon\varepsilon_0}\left(\frac{e^{\kappa R}}{1+\kappa R}\right)^2\frac{e^{-\kappa \rho}}{\rho}.
 \end{equation}
Comparison between Eqs.~\eqref{eq:pair-ICi-Iso} and~\eqref{eq:pair-CSp-Iso} shows that the ICi and the CSp pair interaction energies coincide at all interparticle distances $\rho$ independently of $\kappa R$, i.e.  
\begin{equation}
    \frac{V^\mathrm{ICi}(\rho)}{V^\mathrm{CSp}(\rho)}=1,\quad\textrm{using Eq.~\eqref{eq:qz}}.
\end{equation}
It is worth noting that by simply smearing the ICi charge onto the surface of the CSp model (i.e., by imposing $Q= Z$), one obtains a screening-dependent difference between the pair interaction energies of the two models, namely
\begin{equation}
    \frac{V^\mathrm{ICi}(\rho)}{V^\mathrm{CSp}(\rho)} = \left(\frac{\kappa R}{1+\kappa R}\right)^2\frac{e^{2\kappa R}}{\sinh^2(\kappa R)}\quad\mathrm{when}\quad Q=Z.
\end{equation}


\begin{thebibliography}{47}%
\makeatletter
\providecommand \@ifxundefined [1]{%
 \@ifx{#1\undefined}
}%
\providecommand \@ifnum [1]{%
 \ifnum #1\expandafter \@firstoftwo
 \else \expandafter \@secondoftwo
 \fi
}%
\providecommand \@ifx [1]{%
 \ifx #1\expandafter \@firstoftwo
 \else \expandafter \@secondoftwo
 \fi
}%
\providecommand \natexlab [1]{#1}%
\providecommand \enquote  [1]{``#1''}%
\providecommand \bibnamefont  [1]{#1}%
\providecommand \bibfnamefont [1]{#1}%
\providecommand \citenamefont [1]{#1}%
\providecommand \href@noop [0]{\@secondoftwo}%
\providecommand \href [0]{\begingroup \@sanitize@url \@href}%
\providecommand \@href[1]{\@@startlink{#1}\@@href}%
\providecommand \@@href[1]{\endgroup#1\@@endlink}%
\providecommand \@sanitize@url [0]{\catcode `\\12\catcode `\$12\catcode
  `\&12\catcode `\#12\catcode `\^12\catcode `\_12\catcode `\%12\relax}%
\providecommand \@@startlink[1]{}%
\providecommand \@@endlink[0]{}%
\providecommand \url  [0]{\begingroup\@sanitize@url \@url }%
\providecommand \@url [1]{\endgroup\@href {#1}{\urlprefix }}%
\providecommand \urlprefix  [0]{URL }%
\providecommand \Eprint [0]{\href }%
\providecommand \doibase [0]{https://doi.org/}%
\providecommand \selectlanguage [0]{\@gobble}%
\providecommand \bibinfo  [0]{\@secondoftwo}%
\providecommand \bibfield  [0]{\@secondoftwo}%
\providecommand \translation [1]{[#1]}%
\providecommand \BibitemOpen [0]{}%
\providecommand \bibitemStop [0]{}%
\providecommand \bibitemNoStop [0]{.\EOS\space}%
\providecommand \EOS [0]{\spacefactor3000\relax}%
\providecommand \BibitemShut  [1]{\csname bibitem#1\endcsname}%
\let\auto@bib@innerbib\@empty
\bibitem [{\citenamefont {Yang}\ and\ \citenamefont
  {Rocchia}(2023)}]{yang2023jpcb}%
  \BibitemOpen
  \bibfield  {author} {\bibinfo {author} {\bibfnamefont {W.}~\bibnamefont
  {Yang}}\ and\ \bibinfo {author} {\bibfnamefont {W.}~\bibnamefont {Rocchia}},\
  }\bibfield  {title} {\bibinfo {title} {Biomolecular electrostatic phenomena:
  An evergreen field},\ }\href@noop {} {\bibfield  {journal} {\bibinfo
  {journal} {J. Phys. Chem. B}\ }\textbf {\bibinfo {volume} {127}},\ \bibinfo
  {pages} {3979} (\bibinfo {year} {2023})}\BibitemShut {NoStop}%
\bibitem [{\citenamefont {Hueckel}\ \emph {et~al.}(2021)\citenamefont
  {Hueckel}, \citenamefont {Hocky},\ and\ \citenamefont
  {Sacanna}}]{hueckel2021total}%
  \BibitemOpen
  \bibfield  {author} {\bibinfo {author} {\bibfnamefont {T.}~\bibnamefont
  {Hueckel}}, \bibinfo {author} {\bibfnamefont {G.~M.}\ \bibnamefont {Hocky}},\
  and\ \bibinfo {author} {\bibfnamefont {S.}~\bibnamefont {Sacanna}},\
  }\bibfield  {title} {\bibinfo {title} {Total synthesis of colloidal matter},\
  }\href@noop {} {\bibfield  {journal} {\bibinfo  {journal} {Nat. Rev. Mater.}\
  }\textbf {\bibinfo {volume} {6}},\ \bibinfo {pages} {1053} (\bibinfo {year}
  {2021})}\BibitemShut {NoStop}%
\bibitem [{\citenamefont {Zhou}\ and\ \citenamefont
  {Pang}(2018)}]{zhou2018electrostatic}%
  \BibitemOpen
  \bibfield  {author} {\bibinfo {author} {\bibfnamefont {H.-X.}\ \bibnamefont
  {Zhou}}\ and\ \bibinfo {author} {\bibfnamefont {X.}~\bibnamefont {Pang}},\
  }\bibfield  {title} {\bibinfo {title} {Electrostatic interactions in protein
  structure, folding, binding, and condensation},\ }\href@noop {} {\bibfield
  {journal} {\bibinfo  {journal} {Chem. Rev.}\ }\textbf {\bibinfo {volume}
  {118}},\ \bibinfo {pages} {1691} (\bibinfo {year} {2018})}\BibitemShut
  {NoStop}%
\bibitem [{\citenamefont {van Oostrum}\ \emph {et~al.}(2015)\citenamefont {van
  Oostrum}, \citenamefont {Hejazifar}, \citenamefont {Niedermayer},\ and\
  \citenamefont {Reimhult}}]{vanostrum2015jpcm}%
  \BibitemOpen
  \bibfield  {author} {\bibinfo {author} {\bibfnamefont {P.~D.~J.}\
  \bibnamefont {van Oostrum}}, \bibinfo {author} {\bibfnamefont
  {M.}~\bibnamefont {Hejazifar}}, \bibinfo {author} {\bibfnamefont
  {C.}~\bibnamefont {Niedermayer}},\ and\ \bibinfo {author} {\bibfnamefont
  {E.}~\bibnamefont {Reimhult}},\ }\bibfield  {title} {\bibinfo {title} {Simple
  method for the synthesis of inverse patchy colloids},\ }\href@noop {}
  {\bibfield  {journal} {\bibinfo  {journal} {J. Phys. Condens. Mat.}\ }\textbf
  {\bibinfo {volume} {27}},\ \bibinfo {pages} {234105} (\bibinfo {year}
  {2015})}\BibitemShut {NoStop}%
\bibitem [{\citenamefont {Sabapathy}\ \emph {et~al.}(2017)\citenamefont
  {Sabapathy}, \citenamefont {Mathews},\ and\ \citenamefont
  {Mani}}]{sabapathy2017pccp}%
  \BibitemOpen
  \bibfield  {author} {\bibinfo {author} {\bibfnamefont {M.}~\bibnamefont
  {Sabapathy}}, \bibinfo {author} {\bibfnamefont {R.~A.}\ \bibnamefont
  {Mathews}},\ and\ \bibinfo {author} {\bibfnamefont {E.}~\bibnamefont
  {Mani}},\ }\bibfield  {title} {\bibinfo {title} {Self-assembly of inverse
  patchy colloids with tunable patch coverage},\ }\href@noop {} {\bibfield
  {journal} {\bibinfo  {journal} {Phys. Chem. Chem. Phys.}\ }\textbf {\bibinfo
  {volume} {19}},\ \bibinfo {pages} {13122} (\bibinfo {year}
  {2017})}\BibitemShut {NoStop}%
\bibitem [{\citenamefont {Zimmermann}\ \emph {et~al.}(2018)\citenamefont
  {Zimmermann}, \citenamefont {Grigoriev}, \citenamefont {Puretskiy},\ and\
  \citenamefont {B{\"o}ker}}]{Zimmermann_2018}%
  \BibitemOpen
  \bibfield  {author} {\bibinfo {author} {\bibfnamefont {M.}~\bibnamefont
  {Zimmermann}}, \bibinfo {author} {\bibfnamefont {D.}~\bibnamefont
  {Grigoriev}}, \bibinfo {author} {\bibfnamefont {N.}~\bibnamefont
  {Puretskiy}},\ and\ \bibinfo {author} {\bibfnamefont {A.}~\bibnamefont
  {B{\"o}ker}},\ }\bibfield  {title} {\bibinfo {title} {Characteristics of
  microcontact printing with polyelectrolyte ink for the precise preparation of
  patches on silica particles},\ }\href@noop {} {\bibfield  {journal} {\bibinfo
   {journal} {RSC Adv.}\ }\textbf {\bibinfo {volume} {8}},\ \bibinfo {pages}
  {39241} (\bibinfo {year} {2018})}\BibitemShut {NoStop}%
\bibitem [{\citenamefont {Kierulf}\ \emph {et~al.}(2022)\citenamefont
  {Kierulf}, \citenamefont {Enayati}, \citenamefont {Yaghoobi}, \citenamefont
  {Whaley}, \citenamefont {Smoot}, \citenamefont {Perez~Herrera},\ and\
  \citenamefont {Abbaspourrad}}]{kierulf2022starch}%
  \BibitemOpen
  \bibfield  {author} {\bibinfo {author} {\bibfnamefont {A.}~\bibnamefont
  {Kierulf}}, \bibinfo {author} {\bibfnamefont {M.}~\bibnamefont {Enayati}},
  \bibinfo {author} {\bibfnamefont {M.}~\bibnamefont {Yaghoobi}}, \bibinfo
  {author} {\bibfnamefont {J.}~\bibnamefont {Whaley}}, \bibinfo {author}
  {\bibfnamefont {J.}~\bibnamefont {Smoot}}, \bibinfo {author} {\bibfnamefont
  {M.}~\bibnamefont {Perez~Herrera}},\ and\ \bibinfo {author} {\bibfnamefont
  {A.}~\bibnamefont {Abbaspourrad}},\ }\bibfield  {title} {\bibinfo {title}
  {Starch {Janus} particles: Bulk synthesis, self-assembly, rheology, and
  potential food applications},\ }\href@noop {} {\bibfield  {journal} {\bibinfo
   {journal} {ACS Appl. Mater. Interfaces.}\ }\textbf {\bibinfo {volume}
  {14}},\ \bibinfo {pages} {57371} (\bibinfo {year} {2022})}\BibitemShut
  {NoStop}%
\bibitem [{\citenamefont {Virk}\ \emph {et~al.}(2023)\citenamefont {Virk},
  \citenamefont {Beitl},\ and\ \citenamefont {van Oostrum}}]{Virk2023jpcm}%
  \BibitemOpen
  \bibfield  {author} {\bibinfo {author} {\bibfnamefont {M.~M.}\ \bibnamefont
  {Virk}}, \bibinfo {author} {\bibfnamefont {K.~N.}\ \bibnamefont {Beitl}},\
  and\ \bibinfo {author} {\bibfnamefont {P.~D.~J.}\ \bibnamefont {van
  Oostrum}},\ }\bibfield  {title} {\bibinfo {title} {Synthesis of patchy
  particles using gaseous ligands},\ }\href@noop {} {\bibfield  {journal}
  {\bibinfo  {journal} {J. Phys. Condens. Mat.}\ }\textbf {\bibinfo {volume}
  {35}},\ \bibinfo {pages} {174003} (\bibinfo {year} {2023})}\BibitemShut
  {NoStop}%
\bibitem [{\citenamefont {Li}\ \emph {et~al.}(2015)\citenamefont {Li},
  \citenamefont {Persson}, \citenamefont {Morin}, \citenamefont {Behrens},
  \citenamefont {Lund},\ and\ \citenamefont {Oskolkova}}]{Persson2015jpcb}%
  \BibitemOpen
  \bibfield  {author} {\bibinfo {author} {\bibfnamefont {W.}~\bibnamefont
  {Li}}, \bibinfo {author} {\bibfnamefont {B.~A.}\ \bibnamefont {Persson}},
  \bibinfo {author} {\bibfnamefont {M.}~\bibnamefont {Morin}}, \bibinfo
  {author} {\bibfnamefont {M.~A.}\ \bibnamefont {Behrens}}, \bibinfo {author}
  {\bibfnamefont {M.}~\bibnamefont {Lund}},\ and\ \bibinfo {author}
  {\bibfnamefont {M.~Z.}\ \bibnamefont {Oskolkova}},\ }\bibfield  {title}
  {\bibinfo {title} {Charge-induced patchy attractions between proteins},\
  }\href@noop {} {\bibfield  {journal} {\bibinfo  {journal} {J. Phys. Chem. B}\
  }\textbf {\bibinfo {volume} {119}},\ \bibinfo {pages} {503} (\bibinfo {year}
  {2015})}\BibitemShut {NoStop}%
\bibitem [{\citenamefont {Kress}\ and\ \citenamefont
  {Jones}(2020)}]{kress2020colloidal}%
  \BibitemOpen
  \bibfield  {author} {\bibinfo {author} {\bibfnamefont {R.~N.}\ \bibnamefont
  {Kress}}\ and\ \bibinfo {author} {\bibfnamefont {M.~R.}\ \bibnamefont
  {Jones}},\ }\bibfield  {title} {\bibinfo {title} {Colloidal interactions get
  patchy and directional},\ }\href@noop {} {\bibfield  {journal} {\bibinfo
  {journal} {Proceedings of the National Academy of Sciences}\ }\textbf
  {\bibinfo {volume} {117}},\ \bibinfo {pages} {15382} (\bibinfo {year}
  {2020})}\BibitemShut {NoStop}%
\bibitem [{\citenamefont {Adar}\ \emph {et~al.}(2017)\citenamefont {Adar},
  \citenamefont {Andelman},\ and\ \citenamefont
  {Diamant}}]{adar2017electrostatics}%
  \BibitemOpen
  \bibfield  {author} {\bibinfo {author} {\bibfnamefont {R.~M.}\ \bibnamefont
  {Adar}}, \bibinfo {author} {\bibfnamefont {D.}~\bibnamefont {Andelman}},\
  and\ \bibinfo {author} {\bibfnamefont {H.}~\bibnamefont {Diamant}},\
  }\bibfield  {title} {\bibinfo {title} {Electrostatics of patchy surfaces},\
  }\href@noop {} {\bibfield  {journal} {\bibinfo  {journal} {Adv. Colloid
  Interface Sci.}\ }\textbf {\bibinfo {volume} {247}},\ \bibinfo {pages} {198}
  (\bibinfo {year} {2017})}\BibitemShut {NoStop}%
\bibitem [{\citenamefont {Lunkad}\ \emph {et~al.}(2022)\citenamefont {Lunkad},
  \citenamefont {Barroso~da Silva},\ and\ \citenamefont
  {Kosovan}}]{lunkad2022both}%
  \BibitemOpen
  \bibfield  {author} {\bibinfo {author} {\bibfnamefont {R.}~\bibnamefont
  {Lunkad}}, \bibinfo {author} {\bibfnamefont {F.~L.}\ \bibnamefont {Barroso~da
  Silva}},\ and\ \bibinfo {author} {\bibfnamefont {P.}~\bibnamefont
  {Kosovan}},\ }\bibfield  {title} {\bibinfo {title} {Both charge-regulation
  and charge-patch distribution can drive adsorption on the wrong side of the
  isoelectric point},\ }\href@noop {} {\bibfield  {journal} {\bibinfo
  {journal} {J. Am. Chem. Soc.}\ }\textbf {\bibinfo {volume} {144}},\ \bibinfo
  {pages} {1813} (\bibinfo {year} {2022})}\BibitemShut {NoStop}%
\bibitem [{\citenamefont {Kim}\ \emph {et~al.}(2021)\citenamefont {Kim},
  \citenamefont {Kim}, \citenamefont {Jo}, \citenamefont {Pine}, \citenamefont
  {Sacanna},\ and\ \citenamefont {Yi}}]{kim2021patchy}%
  \BibitemOpen
  \bibfield  {author} {\bibinfo {author} {\bibfnamefont {Y.-J.}\ \bibnamefont
  {Kim}}, \bibinfo {author} {\bibfnamefont {J.-H.}\ \bibnamefont {Kim}},
  \bibinfo {author} {\bibfnamefont {I.-S.}\ \bibnamefont {Jo}}, \bibinfo
  {author} {\bibfnamefont {D.~J.}\ \bibnamefont {Pine}}, \bibinfo {author}
  {\bibfnamefont {S.}~\bibnamefont {Sacanna}},\ and\ \bibinfo {author}
  {\bibfnamefont {G.-R.}\ \bibnamefont {Yi}},\ }\bibfield  {title} {\bibinfo
  {title} {Patchy colloidal clusters with broken symmetry},\ }\href@noop {}
  {\bibfield  {journal} {\bibinfo  {journal} {Journal of the American Chemical
  Society}\ }\textbf {\bibinfo {volume} {143}},\ \bibinfo {pages} {13175}
  (\bibinfo {year} {2021})}\BibitemShut {NoStop}%
\bibitem [{\citenamefont {Noguchi}\ \emph {et~al.}(2019)\citenamefont
  {Noguchi}, \citenamefont {Iwashita},\ and\ \citenamefont
  {Kimura}}]{Kimura_2019}%
  \BibitemOpen
  \bibfield  {author} {\bibinfo {author} {\bibfnamefont {T.~G.}\ \bibnamefont
  {Noguchi}}, \bibinfo {author} {\bibfnamefont {Y.}~\bibnamefont {Iwashita}},\
  and\ \bibinfo {author} {\bibfnamefont {Y.}~\bibnamefont {Kimura}},\
  }\bibfield  {title} {\bibinfo {title} {Controlled armoring of metal surfaces
  with metallodielectric patchy particles},\ }\href@noop {} {\bibfield
  {journal} {\bibinfo  {journal} {J. Chem. Phys.}\ }\textbf {\bibinfo {volume}
  {150}},\ \bibinfo {pages} {174903} (\bibinfo {year} {2019})}\BibitemShut
  {NoStop}%
\bibitem [{\citenamefont {Lebdioua}\ \emph {et~al.}(2021)\citenamefont
  {Lebdioua}, \citenamefont {Cerbelaud}, \citenamefont {Aimable},\ and\
  \citenamefont {Videcoq}}]{lebdioua2021jcis}%
  \BibitemOpen
  \bibfield  {author} {\bibinfo {author} {\bibfnamefont {K.}~\bibnamefont
  {Lebdioua}}, \bibinfo {author} {\bibfnamefont {M.}~\bibnamefont {Cerbelaud}},
  \bibinfo {author} {\bibfnamefont {A.}~\bibnamefont {Aimable}},\ and\ \bibinfo
  {author} {\bibfnamefont {A.}~\bibnamefont {Videcoq}},\ }\bibfield  {title}
  {\bibinfo {title} {Study of the aggregation behavior of {Janus} particles by
  coupling experiments and {Brownian} dynamics simulations},\ }\href@noop {}
  {\bibfield  {journal} {\bibinfo  {journal} {J. Colloid Interface Sci.}\
  }\textbf {\bibinfo {volume} {583}},\ \bibinfo {pages} {222} (\bibinfo {year}
  {2021})}\BibitemShut {NoStop}%
\bibitem [{\citenamefont {Naderi~Mehr}\ \emph {et~al.}(2020)\citenamefont
  {Naderi~Mehr}, \citenamefont {Grigoriev}, \citenamefont {Heaton},
  \citenamefont {Baptiste}, \citenamefont {Stace}, \citenamefont {Puretskiy},
  \citenamefont {Besley},\ and\ \citenamefont {B{\"o}ker}}]{naderi2020self}%
  \BibitemOpen
  \bibfield  {author} {\bibinfo {author} {\bibfnamefont {F.}~\bibnamefont
  {Naderi~Mehr}}, \bibinfo {author} {\bibfnamefont {D.}~\bibnamefont
  {Grigoriev}}, \bibinfo {author} {\bibfnamefont {R.}~\bibnamefont {Heaton}},
  \bibinfo {author} {\bibfnamefont {J.}~\bibnamefont {Baptiste}}, \bibinfo
  {author} {\bibfnamefont {A.~J.}\ \bibnamefont {Stace}}, \bibinfo {author}
  {\bibfnamefont {N.}~\bibnamefont {Puretskiy}}, \bibinfo {author}
  {\bibfnamefont {E.}~\bibnamefont {Besley}},\ and\ \bibinfo {author}
  {\bibfnamefont {A.}~\bibnamefont {B{\"o}ker}},\ }\bibfield  {title} {\bibinfo
  {title} {Self-assembly behavior of oppositely charged inverse bipatchy
  microcolloids},\ }\href@noop {} {\bibfield  {journal} {\bibinfo  {journal}
  {Small}\ }\textbf {\bibinfo {volume} {16}},\ \bibinfo {pages} {2000442}
  (\bibinfo {year} {2020})}\BibitemShut {NoStop}%
\bibitem [{\citenamefont {Li}\ \emph {et~al.}(2022)\citenamefont {Li},
  \citenamefont {Kierulf}, \citenamefont {Whaley}, \citenamefont {Smoot},
  \citenamefont {Herrera},\ and\ \citenamefont
  {Abbaspourrad}}]{li2022modulating}%
  \BibitemOpen
  \bibfield  {author} {\bibinfo {author} {\bibfnamefont {P.}~\bibnamefont
  {Li}}, \bibinfo {author} {\bibfnamefont {A.}~\bibnamefont {Kierulf}},
  \bibinfo {author} {\bibfnamefont {J.}~\bibnamefont {Whaley}}, \bibinfo
  {author} {\bibfnamefont {J.}~\bibnamefont {Smoot}}, \bibinfo {author}
  {\bibfnamefont {M.~P.}\ \bibnamefont {Herrera}},\ and\ \bibinfo {author}
  {\bibfnamefont {A.}~\bibnamefont {Abbaspourrad}},\ }\bibfield  {title}
  {\bibinfo {title} {Modulating functionality of starch-based patchy particles
  by manipulating architecture and environmental factors},\ }\href@noop {}
  {\bibfield  {journal} {\bibinfo  {journal} {ACS Appl. Mater. Interfaces.}\
  }\textbf {\bibinfo {volume} {14}},\ \bibinfo {pages} {39497} (\bibinfo {year}
  {2022})}\BibitemShut {NoStop}%
\bibitem [{\citenamefont {Guo}\ \emph {et~al.}(2021)\citenamefont {Guo},
  \citenamefont {Nishida},\ and\ \citenamefont {Hoshino}}]{guo2021quantifying}%
  \BibitemOpen
  \bibfield  {author} {\bibinfo {author} {\bibfnamefont {Y.}~\bibnamefont
  {Guo}}, \bibinfo {author} {\bibfnamefont {N.}~\bibnamefont {Nishida}},\ and\
  \bibinfo {author} {\bibfnamefont {T.}~\bibnamefont {Hoshino}},\ }\bibfield
  {title} {\bibinfo {title} {Quantifying the separation of positive and
  negative areas in electrostatic potential for predicting feasibility of
  ammonium sulfate for protein crystallization},\ }\href@noop {} {\bibfield
  {journal} {\bibinfo  {journal} {J. Chem. Inf. Model.}\ }\textbf {\bibinfo
  {volume} {61}},\ \bibinfo {pages} {4571} (\bibinfo {year}
  {2021})}\BibitemShut {NoStop}%
\bibitem [{\citenamefont {Zhang}\ \emph {et~al.}(2020)\citenamefont {Zhang},
  \citenamefont {Alberstein}, \citenamefont {De~Yoreo},\ and\ \citenamefont
  {Tezcan}}]{zhang2020assembly}%
  \BibitemOpen
  \bibfield  {author} {\bibinfo {author} {\bibfnamefont {S.}~\bibnamefont
  {Zhang}}, \bibinfo {author} {\bibfnamefont {R.~G.}\ \bibnamefont
  {Alberstein}}, \bibinfo {author} {\bibfnamefont {J.~J.}\ \bibnamefont
  {De~Yoreo}},\ and\ \bibinfo {author} {\bibfnamefont {F.~A.}\ \bibnamefont
  {Tezcan}},\ }\bibfield  {title} {\bibinfo {title} {Assembly of a patchy
  protein into variable {2D} lattices via tunable multiscale interactions},\
  }\href@noop {} {\bibfield  {journal} {\bibinfo  {journal} {Nat. Commun.}\
  }\textbf {\bibinfo {volume} {11}},\ \bibinfo {pages} {3770} (\bibinfo {year}
  {2020})}\BibitemShut {NoStop}%
\bibitem [{\citenamefont {Ausserw{\"o}ger}\ \emph {et~al.}(2023)\citenamefont
  {Ausserw{\"o}ger}, \citenamefont {Krainer}, \citenamefont {Welsh},
  \citenamefont {Thorsteinson}, \citenamefont {{de Csill{\'e}ry}},
  \citenamefont {Sneideris}, \citenamefont {Schneider}, \citenamefont
  {Egebjerg}, \citenamefont {Invernizzi}, \citenamefont {Herling},
  \citenamefont {Lorenzen},\ and\ \citenamefont {Knowles}}]{Knowles_2023}%
  \BibitemOpen
  \bibfield  {author} {\bibinfo {author} {\bibfnamefont {H.}~\bibnamefont
  {Ausserw{\"o}ger}}, \bibinfo {author} {\bibfnamefont {G.}~\bibnamefont
  {Krainer}}, \bibinfo {author} {\bibfnamefont {T.~J.}\ \bibnamefont {Welsh}},
  \bibinfo {author} {\bibfnamefont {N.}~\bibnamefont {Thorsteinson}}, \bibinfo
  {author} {\bibfnamefont {E.}~\bibnamefont {{de Csill{\'e}ry}}}, \bibinfo
  {author} {\bibfnamefont {T.}~\bibnamefont {Sneideris}}, \bibinfo {author}
  {\bibfnamefont {M.~M.}\ \bibnamefont {Schneider}}, \bibinfo {author}
  {\bibfnamefont {T.}~\bibnamefont {Egebjerg}}, \bibinfo {author}
  {\bibfnamefont {G.}~\bibnamefont {Invernizzi}}, \bibinfo {author}
  {\bibfnamefont {T.~W.}\ \bibnamefont {Herling}}, \bibinfo {author}
  {\bibfnamefont {N.}~\bibnamefont {Lorenzen}},\ and\ \bibinfo {author}
  {\bibfnamefont {T.~P.~J.}\ \bibnamefont {Knowles}},\ }\bibfield  {title}
  {\bibinfo {title} {Surface patches induce nonspecific binding and phase
  separation of antibodies},\ }\href@noop {} {\bibfield  {journal} {\bibinfo
  {journal} {PNAS}\ }\textbf {\bibinfo {volume} {120}},\ \bibinfo {pages}
  {e2210332120} (\bibinfo {year} {2023})}\BibitemShut {NoStop}%
\bibitem [{\citenamefont {Kim}\ \emph {et~al.}(2024)\citenamefont {Kim},
  \citenamefont {Qin}, \citenamefont {Zhou},\ and\ \citenamefont
  {Rosen}}]{kim2024surface}%
  \BibitemOpen
  \bibfield  {author} {\bibinfo {author} {\bibfnamefont {J.}~\bibnamefont
  {Kim}}, \bibinfo {author} {\bibfnamefont {S.}~\bibnamefont {Qin}}, \bibinfo
  {author} {\bibfnamefont {H.-X.}\ \bibnamefont {Zhou}},\ and\ \bibinfo
  {author} {\bibfnamefont {M.~K.}\ \bibnamefont {Rosen}},\ }\bibfield  {title}
  {\bibinfo {title} {Surface charge can modulate phase separation of
  multidomain proteins},\ }\href@noop {} {\bibfield  {journal} {\bibinfo
  {journal} {J. Am. Chem. Soc.}\ } (\bibinfo {year} {2024})}\BibitemShut
  {NoStop}%
\bibitem [{\citenamefont {Besley}(2023)}]{besley2023recent}%
  \BibitemOpen
  \bibfield  {author} {\bibinfo {author} {\bibfnamefont {E.}~\bibnamefont
  {Besley}},\ }\bibfield  {title} {\bibinfo {title} {Recent developments in the
  methods and applications of electrostatic theory},\ }\href@noop {} {\bibfield
   {journal} {\bibinfo  {journal} {Acc. Chem. Res.}\ }\textbf {\bibinfo
  {volume} {56}},\ \bibinfo {pages} {2267} (\bibinfo {year}
  {2023})}\BibitemShut {NoStop}%
\bibitem [{\citenamefont {Siryk}\ \emph {et~al.}(2021)\citenamefont {Siryk},
  \citenamefont {Bendandi}, \citenamefont {Diaspro},\ and\ \citenamefont
  {Rocchia}}]{siryk2021jcp}%
  \BibitemOpen
  \bibfield  {author} {\bibinfo {author} {\bibfnamefont {S.~V.}\ \bibnamefont
  {Siryk}}, \bibinfo {author} {\bibfnamefont {A.}~\bibnamefont {Bendandi}},
  \bibinfo {author} {\bibfnamefont {A.}~\bibnamefont {Diaspro}},\ and\ \bibinfo
  {author} {\bibfnamefont {W.}~\bibnamefont {Rocchia}},\ }\bibfield  {title}
  {\bibinfo {title} {Charged dielectric spheres interacting in electrolytic
  solution: A linearized {Poisson--Boltzmann} equation model},\ }\href@noop {}
  {\bibfield  {journal} {\bibinfo  {journal} {J. Chem. Phys.}\ }\textbf
  {\bibinfo {volume} {155}},\ \bibinfo {pages} {114114} (\bibinfo {year}
  {2021})}\BibitemShut {NoStop}%
\bibitem [{\citenamefont {{de Graaf}}\ \emph {et~al.}(2012)\citenamefont {{de
  Graaf}}, \citenamefont {Boon}, \citenamefont {Dijkstra},\ and\ \citenamefont
  {{van Roij}}}]{Boon2012}%
  \BibitemOpen
  \bibfield  {author} {\bibinfo {author} {\bibfnamefont {J.}~\bibnamefont {{de
  Graaf}}}, \bibinfo {author} {\bibfnamefont {N.}~\bibnamefont {Boon}},
  \bibinfo {author} {\bibfnamefont {M.}~\bibnamefont {Dijkstra}},\ and\
  \bibinfo {author} {\bibfnamefont {R.}~\bibnamefont {{van Roij}}},\ }\bibfield
   {title} {\bibinfo {title} {Electrostatic interactions between {Janus}
  particles},\ }\href@noop {} {\bibfield  {journal} {\bibinfo  {journal} {The
  Journal of chemical physics}\ }\textbf {\bibinfo {volume} {137}},\ \bibinfo
  {pages} {104910} (\bibinfo {year} {2012})}\BibitemShut {NoStop}%
\bibitem [{\citenamefont {Bo\v{z}i\v{c}}\ and\ \citenamefont
  {Podgornik}(2013)}]{bozic2013jcp}%
  \BibitemOpen
  \bibfield  {author} {\bibinfo {author} {\bibfnamefont {A.}~\bibnamefont
  {Bo\v{z}i\v{c}}}\ and\ \bibinfo {author} {\bibfnamefont {R.}~\bibnamefont
  {Podgornik}},\ }\bibfield  {title} {\bibinfo {title} {Symmetry effects in
  electrostatic interactions between two arbitrarily charged spherical shells
  in the {D}ebye-{H}{\"u}ckel approximation},\ }\href@noop {} {\bibfield
  {journal} {\bibinfo  {journal} {J. Chem. Phys.}\ }\textbf {\bibinfo {volume}
  {138}},\ \bibinfo {pages} {074902} (\bibinfo {year} {2013})}\BibitemShut
  {NoStop}%
\bibitem [{\citenamefont {Obolensky}\ \emph {et~al.}(2021)\citenamefont
  {Obolensky}, \citenamefont {Doerr},\ and\ \citenamefont
  {Yu}}]{obolensky2021rigorous}%
  \BibitemOpen
  \bibfield  {author} {\bibinfo {author} {\bibfnamefont {O.}~\bibnamefont
  {Obolensky}}, \bibinfo {author} {\bibfnamefont {T.}~\bibnamefont {Doerr}},\
  and\ \bibinfo {author} {\bibfnamefont {Y.-K.}\ \bibnamefont {Yu}},\
  }\bibfield  {title} {\bibinfo {title} {Rigorous treatment of pairwise and
  many-body electrostatic interactions among dielectric spheres at the
  {Debye--H{\"u}ckel} level},\ }\href@noop {} {\bibfield  {journal} {\bibinfo
  {journal} {Eur. Phys. J. E}\ }\textbf {\bibinfo {volume} {44}},\ \bibinfo
  {pages} {1} (\bibinfo {year} {2021})}\BibitemShut {NoStop}%
\bibitem [{\citenamefont {Bianchi}\ \emph {et~al.}(2011)\citenamefont
  {Bianchi}, \citenamefont {Kahl},\ and\ \citenamefont
  {Likos}}]{bianchi2011sm}%
  \BibitemOpen
  \bibfield  {author} {\bibinfo {author} {\bibfnamefont {E.}~\bibnamefont
  {Bianchi}}, \bibinfo {author} {\bibfnamefont {G.}~\bibnamefont {Kahl}},\ and\
  \bibinfo {author} {\bibfnamefont {C.~N.}\ \bibnamefont {Likos}},\ }\bibfield
  {title} {\bibinfo {title} {Inverse patchy colloids: from microscopic
  description to mesoscopic coarse-graining},\ }\href@noop {} {\bibfield
  {journal} {\bibinfo  {journal} {Soft Matter}\ }\textbf {\bibinfo {volume}
  {7}},\ \bibinfo {pages} {8313} (\bibinfo {year} {2011})}\BibitemShut
  {NoStop}%
\bibitem [{\citenamefont {Stipsitz}\ \emph {et~al.}(2015)\citenamefont
  {Stipsitz}, \citenamefont {Bianchi},\ and\ \citenamefont
  {Kahl}}]{bianchi:2015}%
  \BibitemOpen
  \bibfield  {author} {\bibinfo {author} {\bibfnamefont {M.}~\bibnamefont
  {Stipsitz}}, \bibinfo {author} {\bibfnamefont {E.}~\bibnamefont {Bianchi}},\
  and\ \bibinfo {author} {\bibfnamefont {G.}~\bibnamefont {Kahl}},\ }\bibfield
  {title} {\bibinfo {title} {Generalized inverse patchy colloid model},\
  }\href@noop {} {\bibfield  {journal} {\bibinfo  {journal} {J. Chem. Phys.}\
  }\textbf {\bibinfo {volume} {142}},\ \bibinfo {pages} {114905} (\bibinfo
  {year} {2015})}\BibitemShut {NoStop}%
\bibitem [{\citenamefont {Wu}\ \emph {et~al.}(2016)\citenamefont {Wu},
  \citenamefont {Han},\ and\ \citenamefont {Luijten}}]{Luijten2016}%
  \BibitemOpen
  \bibfield  {author} {\bibinfo {author} {\bibfnamefont {H.}~\bibnamefont
  {Wu}}, \bibinfo {author} {\bibfnamefont {M.}~\bibnamefont {Han}},\ and\
  \bibinfo {author} {\bibfnamefont {E.}~\bibnamefont {Luijten}},\ }\bibfield
  {title} {\bibinfo {title} {Dielectric effects on the ion distribution near a
  {Janus} colloid},\ }\href@noop {} {\bibfield  {journal} {\bibinfo  {journal}
  {Soft Matter}\ }\textbf {\bibinfo {volume} {12}},\ \bibinfo {pages} {9575}
  (\bibinfo {year} {2016})}\BibitemShut {NoStop}%
\bibitem [{\citenamefont {Popov}\ and\ \citenamefont
  {Hernandez}(2023)}]{popov2023jpcb}%
  \BibitemOpen
  \bibfield  {author} {\bibinfo {author} {\bibfnamefont {A.}~\bibnamefont
  {Popov}}\ and\ \bibinfo {author} {\bibfnamefont {R.}~\bibnamefont
  {Hernandez}},\ }\bibfield  {title} {\bibinfo {title} {Bottom-up construction
  of the interaction between {Janus} particles},\ }\href@noop {} {\bibfield
  {journal} {\bibinfo  {journal} {J. Phys. Chem. B}\ }\textbf {\bibinfo
  {volume} {127}},\ \bibinfo {pages} {1664} (\bibinfo {year}
  {2023})}\BibitemShut {NoStop}%
\bibitem [{\citenamefont {Yigit}\ \emph {et~al.}(2015)\citenamefont {Yigit},
  \citenamefont {Heyda},\ and\ \citenamefont {Dzubiella}}]{yigit15a}%
  \BibitemOpen
  \bibfield  {author} {\bibinfo {author} {\bibfnamefont {C.}~\bibnamefont
  {Yigit}}, \bibinfo {author} {\bibfnamefont {J.}~\bibnamefont {Heyda}},\ and\
  \bibinfo {author} {\bibfnamefont {J.}~\bibnamefont {Dzubiella}},\ }\bibfield
  {title} {\bibinfo {title} {Charged patchy particle models in explicit salt:
  ion distributions, electrostatic potentials, and effective interactions},\
  }\href@noop {} {\bibfield  {journal} {\bibinfo  {journal} {J. Chem. Phys.}\
  }\textbf {\bibinfo {volume} {143}},\ \bibinfo {pages} {064904} (\bibinfo
  {year} {2015})}\BibitemShut {NoStop}%
\bibitem [{\citenamefont {Mathews}\ and\ \citenamefont
  {Mani}(2021)}]{mani2021stabilizing}%
  \BibitemOpen
  \bibfield  {author} {\bibinfo {author} {\bibfnamefont {R.~A.~K.}\
  \bibnamefont {Mathews}}\ and\ \bibinfo {author} {\bibfnamefont
  {E.}~\bibnamefont {Mani}},\ }\bibfield  {title} {\bibinfo {title}
  {Stabilizing ordered structures with single patch inverse patchy colloids in
  two dimensions},\ }\href@noop {} {\bibfield  {journal} {\bibinfo  {journal}
  {J. Phys. Condens. Mat.}\ }\textbf {\bibinfo {volume} {33}},\ \bibinfo
  {pages} {195101} (\bibinfo {year} {2021})}\BibitemShut {NoStop}%
\bibitem [{\citenamefont {Ferrari}\ \emph {et~al.}(2017)\citenamefont
  {Ferrari}, \citenamefont {Bianchi},\ and\ \citenamefont
  {Kahl}}]{silvanonanoscale}%
  \BibitemOpen
  \bibfield  {author} {\bibinfo {author} {\bibfnamefont {S.}~\bibnamefont
  {Ferrari}}, \bibinfo {author} {\bibfnamefont {E.}~\bibnamefont {Bianchi}},\
  and\ \bibinfo {author} {\bibfnamefont {G.}~\bibnamefont {Kahl}},\ }\bibfield
  {title} {\bibinfo {title} {Spontaneous assembly of a hybrid crystal-liquid
  phase in inverse patchy colloid systems},\ }\href@noop {} {\bibfield
  {journal} {\bibinfo  {journal} {Nanoscale}\ }\textbf {\bibinfo {volume}
  {9}},\ \bibinfo {pages} {1956} (\bibinfo {year} {2017})}\BibitemShut
  {NoStop}%
\bibitem [{\citenamefont {Wang}\ and\ \citenamefont {Swan}(2019)}]{Swan_2019}%
  \BibitemOpen
  \bibfield  {author} {\bibinfo {author} {\bibfnamefont {G.}~\bibnamefont
  {Wang}}\ and\ \bibinfo {author} {\bibfnamefont {J.~W.}\ \bibnamefont
  {Swan}},\ }\bibfield  {title} {\bibinfo {title} {Surface heterogeneity
  affects percolation and gelation of colloids: dynamic simulations with random
  patchy spheres},\ }\href@noop {} {\bibfield  {journal} {\bibinfo  {journal}
  {Soft Matter}\ }\textbf {\bibinfo {volume} {215}},\ \bibinfo {pages} {5094}
  (\bibinfo {year} {2019})}\BibitemShut {NoStop}%
\bibitem [{\citenamefont {Yigit}\ \emph {et~al.}(2017)\citenamefont {Yigit},
  \citenamefont {Kandu{\v c}}, \citenamefont {Ballauff},\ and\ \citenamefont
  {Dzubiella}}]{yigit17}%
  \BibitemOpen
  \bibfield  {author} {\bibinfo {author} {\bibfnamefont {C.}~\bibnamefont
  {Yigit}}, \bibinfo {author} {\bibfnamefont {M.}~\bibnamefont {Kandu{\v c}}},
  \bibinfo {author} {\bibfnamefont {M.}~\bibnamefont {Ballauff}},\ and\
  \bibinfo {author} {\bibfnamefont {J.}~\bibnamefont {Dzubiella}},\ }\bibfield
  {title} {\bibinfo {title} {Interaction of charged patchy protein models with
  like-charged polyelectrolyte brushes},\ }\href@noop {} {\bibfield  {journal}
  {\bibinfo  {journal} {Langmuir}\ }\textbf {\bibinfo {volume} {33}},\ \bibinfo
  {pages} {417} (\bibinfo {year} {2017})}\BibitemShut {NoStop}%
\bibitem [{\citenamefont {Verwey}\ and\ \citenamefont
  {Overbeek}(1948)}]{VerweyOverbeek-1948}%
  \BibitemOpen
  \bibfield  {author} {\bibinfo {author} {\bibfnamefont {E.~J.~W.}\
  \bibnamefont {Verwey}}\ and\ \bibinfo {author} {\bibfnamefont {J.~T.~G.}\
  \bibnamefont {Overbeek}},\ }\href@noop {} {\emph {\bibinfo {title} {Theory of
  the Stability of Lyophobic Colloids}}}\ (\bibinfo  {publisher} {Elsevier,
  Amsterdam},\ \bibinfo {year} {1948})\BibitemShut {NoStop}%
\bibitem [{\citenamefont {Bo\v{z}i\v{c}}\ \emph {et~al.}(2012)\citenamefont
  {Bo\v{z}i\v{c}}, \citenamefont {\v{S}iber},\ and\ \citenamefont
  {Podgornik}}]{bozic2012jbp}%
  \BibitemOpen
  \bibfield  {author} {\bibinfo {author} {\bibfnamefont {A.}~\bibnamefont
  {Bo\v{z}i\v{c}}}, \bibinfo {author} {\bibfnamefont {A.}~\bibnamefont
  {\v{S}iber}},\ and\ \bibinfo {author} {\bibfnamefont {R.}~\bibnamefont
  {Podgornik}},\ }\bibfield  {title} {\bibinfo {title} {How simple can a model
  of an empty viral capsid be? {C}harge distributions in viral capsids},\
  }\href@noop {} {\bibfield  {journal} {\bibinfo  {journal} {J. Biol. Phys.}\
  }\textbf {\bibinfo {volume} {38}},\ \bibinfo {pages} {657} (\bibinfo {year}
  {2012})}\BibitemShut {NoStop}%
\bibitem [{\citenamefont {Everts}(2020)}]{everts2020screened}%
  \BibitemOpen
  \bibfield  {author} {\bibinfo {author} {\bibfnamefont {J.~C.}\ \bibnamefont
  {Everts}},\ }\bibfield  {title} {\bibinfo {title} {Screened coulomb
  interactions of general macroions with nonzero particle volume},\ }\href@noop
  {} {\bibfield  {journal} {\bibinfo  {journal} {Phys. Rev. Res.}\ }\textbf
  {\bibinfo {volume} {2}},\ \bibinfo {pages} {033144} (\bibinfo {year}
  {2020})}\BibitemShut {NoStop}%
\bibitem [{\citenamefont {Hong}\ \emph {et~al.}(2006)\citenamefont {Hong},
  \citenamefont {Cacciuto}, \citenamefont {Luijten},\ and\ \citenamefont
  {Granick}}]{hong2006clusters}%
  \BibitemOpen
  \bibfield  {author} {\bibinfo {author} {\bibfnamefont {L.}~\bibnamefont
  {Hong}}, \bibinfo {author} {\bibfnamefont {A.}~\bibnamefont {Cacciuto}},
  \bibinfo {author} {\bibfnamefont {E.}~\bibnamefont {Luijten}},\ and\ \bibinfo
  {author} {\bibfnamefont {S.}~\bibnamefont {Granick}},\ }\bibfield  {title}
  {\bibinfo {title} {Clusters of charged {Janus} spheres},\ }\href@noop {}
  {\bibfield  {journal} {\bibinfo  {journal} {Nano Lett.}\ }\textbf {\bibinfo
  {volume} {6}},\ \bibinfo {pages} {2510} (\bibinfo {year} {2006})}\BibitemShut
  {NoStop}%
\bibitem [{\citenamefont {Hoppe}(2013)}]{hoppe2013simplified}%
  \BibitemOpen
  \bibfield  {author} {\bibinfo {author} {\bibfnamefont {T.}~\bibnamefont
  {Hoppe}},\ }\bibfield  {title} {\bibinfo {title} {A simplified representation
  of anisotropic charge distributions within proteins},\ }\href@noop {}
  {\bibfield  {journal} {\bibinfo  {journal} {J. Chem. Phys.}\ }\textbf
  {\bibinfo {volume} {138}} (\bibinfo {year} {2013})}\BibitemShut {NoStop}%
\bibitem [{\citenamefont {Blanco}\ and\ \citenamefont
  {Shen}(2016)}]{Blanco_2016}%
  \BibitemOpen
  \bibfield  {author} {\bibinfo {author} {\bibfnamefont {M.~A.}\ \bibnamefont
  {Blanco}}\ and\ \bibinfo {author} {\bibfnamefont {V.~K.}\ \bibnamefont
  {Shen}},\ }\bibfield  {title} {\bibinfo {title} {Effect of the surface charge
  distribution on the fluid phase behavior of charged colloids and proteins},\
  }\href@noop {} {\bibfield  {journal} {\bibinfo  {journal} {J. Chem. Phys.}\
  }\textbf {\bibinfo {volume} {145}},\ \bibinfo {pages} {155102} (\bibinfo
  {year} {2016})}\BibitemShut {NoStop}%
\bibitem [{\citenamefont {Bo{\v{z}}i{\v{c}}}\ and\ \citenamefont
  {Podgornik}(2017)}]{bozic2017ph}%
  \BibitemOpen
  \bibfield  {author} {\bibinfo {author} {\bibfnamefont {A.}~\bibnamefont
  {Bo{\v{z}}i{\v{c}}}}\ and\ \bibinfo {author} {\bibfnamefont {R.}~\bibnamefont
  {Podgornik}},\ }\bibfield  {title} {\bibinfo {title} {{pH} dependence of
  charge multipole moments in proteins},\ }\href@noop {} {\bibfield  {journal}
  {\bibinfo  {journal} {Biophys. J.}\ }\textbf {\bibinfo {volume} {113}},\
  \bibinfo {pages} {1454} (\bibinfo {year} {2017})}\BibitemShut {NoStop}%
\bibitem [{\citenamefont {Trizac}\ \emph {et~al.}(2002)\citenamefont {Trizac},
  \citenamefont {Bocquet},\ and\ \citenamefont {Aubouy}}]{trizac2002simple}%
  \BibitemOpen
  \bibfield  {author} {\bibinfo {author} {\bibfnamefont {E.}~\bibnamefont
  {Trizac}}, \bibinfo {author} {\bibfnamefont {L.}~\bibnamefont {Bocquet}},\
  and\ \bibinfo {author} {\bibfnamefont {M.}~\bibnamefont {Aubouy}},\
  }\bibfield  {title} {\bibinfo {title} {Simple approach for charge
  renormalization in highly charged macroions},\ }\href@noop {} {\bibfield
  {journal} {\bibinfo  {journal} {Phys. Rev. Lett.}\ }\textbf {\bibinfo
  {volume} {89}},\ \bibinfo {pages} {248301} (\bibinfo {year}
  {2002})}\BibitemShut {NoStop}%
\bibitem [{\citenamefont {Trizac}\ \emph {et~al.}(2003)\citenamefont {Trizac},
  \citenamefont {Bocquet}, \citenamefont {Aubouy},\ and\ \citenamefont {von
  Gr{\"u}nberg}}]{trizac2003alexander}%
  \BibitemOpen
  \bibfield  {author} {\bibinfo {author} {\bibfnamefont {E.}~\bibnamefont
  {Trizac}}, \bibinfo {author} {\bibfnamefont {L.}~\bibnamefont {Bocquet}},
  \bibinfo {author} {\bibfnamefont {M.}~\bibnamefont {Aubouy}},\ and\ \bibinfo
  {author} {\bibfnamefont {H.-H.}\ \bibnamefont {von Gr{\"u}nberg}},\
  }\bibfield  {title} {\bibinfo {title} {Alexander's prescription for colloidal
  charge renormalization},\ }\href@noop {} {\bibfield  {journal} {\bibinfo
  {journal} {Langmuir}\ }\textbf {\bibinfo {volume} {19}},\ \bibinfo {pages}
  {4027} (\bibinfo {year} {2003})}\BibitemShut {NoStop}%
\bibitem [{\citenamefont {Bo{\v{z}}i{\v{c}}}\ and\ \citenamefont
  {Podgornik}(2018)}]{bozic2018jcp}%
  \BibitemOpen
  \bibfield  {author} {\bibinfo {author} {\bibfnamefont {A.}~\bibnamefont
  {Bo{\v{z}}i{\v{c}}}}\ and\ \bibinfo {author} {\bibfnamefont {R.}~\bibnamefont
  {Podgornik}},\ }\bibfield  {title} {\bibinfo {title} {Anomalous multipole
  expansion: Charge regulation of patchy inhomogeneously charged spherical
  particles},\ }\href@noop {} {\bibfield  {journal} {\bibinfo  {journal} {J.
  Chem. Phys.}\ }\textbf {\bibinfo {volume} {149}},\ \bibinfo {pages} {163307}
  (\bibinfo {year} {2018})}\BibitemShut {NoStop}%
\bibitem [{\citenamefont {Pusara}\ \emph {et~al.}(2021)\citenamefont {Pusara},
  \citenamefont {Yamin}, \citenamefont {Wenzel}, \citenamefont {Krsti{\'c}},\
  and\ \citenamefont {Kozlowska}}]{pusara2021coarse}%
  \BibitemOpen
  \bibfield  {author} {\bibinfo {author} {\bibfnamefont {S.}~\bibnamefont
  {Pusara}}, \bibinfo {author} {\bibfnamefont {P.}~\bibnamefont {Yamin}},
  \bibinfo {author} {\bibfnamefont {W.}~\bibnamefont {Wenzel}}, \bibinfo
  {author} {\bibfnamefont {M.}~\bibnamefont {Krsti{\'c}}},\ and\ \bibinfo
  {author} {\bibfnamefont {M.}~\bibnamefont {Kozlowska}},\ }\bibfield  {title}
  {\bibinfo {title} {A coarse-grained {xDLVO} model for colloidal
  protein--protein interactions},\ }\href@noop {} {\bibfield  {journal}
  {\bibinfo  {journal} {Phys. Chem. Chem. Phys.}\ }\textbf {\bibinfo {volume}
  {23}},\ \bibinfo {pages} {12780} (\bibinfo {year} {2021})}\BibitemShut
  {NoStop}%
\bibitem [{\citenamefont {Lund}\ and\ \citenamefont
  {J{\"o}nsson}(2005)}]{lund2005charge}%
  \BibitemOpen
  \bibfield  {author} {\bibinfo {author} {\bibfnamefont {M.}~\bibnamefont
  {Lund}}\ and\ \bibinfo {author} {\bibfnamefont {B.}~\bibnamefont
  {J{\"o}nsson}},\ }\bibfield  {title} {\bibinfo {title} {On the charge
  regulation of proteins},\ }\href@noop {} {\bibfield  {journal} {\bibinfo
  {journal} {Biochem.}\ }\textbf {\bibinfo {volume} {44}},\ \bibinfo {pages}
  {5722} (\bibinfo {year} {2005})}\BibitemShut {NoStop}%
\end{thebibliography}

\begin{thebibliography}{1}

\bibitem{bianchi2011sm}
Emanuela Bianchi, Gerhard Kahl, and Christos~N Likos.
\newblock Inverse patchy colloids: from microscopic description to mesoscopic
  coarse-graining.
\newblock {\em Soft Matter}, 7(18):8313--8323, 2011.

\bibitem{bozic2013jcp}
A.~Bo\v{z}i\v{c} and R.~Podgornik.
\newblock Symmetry effects in electrostatic interactions between two
  arbitrarily charged spherical shells in the {D}ebye-{H}{\"u}ckel
  approximation.
\newblock {\em J. Chem. Phys.}, 138:074902, 2013.

\bibitem{bozic2018jcp}
An{\v{z}}e Bo{\v{z}}i{\v{c}} and Rudolf Podgornik.
\newblock Anomalous multipole expansion: Charge regulation of patchy
  inhomogeneously charged spherical particles.
\newblock {\em J. Chem. Phys.}, 149:163307, 2018.

\bibitem{hoffmann2004molphys}
N.~Hoffmann, C.~N. Likos, and J.-P. Hansen.
\newblock Linear screening of the electrostatic potential around spherical
  particles with non-spherical charge patterns.
\newblock {\em Mol. Phys.}, 102:857, 2004.

\bibitem{abramowitz1972}
Milton Abramowitz and Irene~A. Stegun.
\newblock {\em Handbook of Mathematical Functions with Formulas, Graphs, and
  Mathematical Tables}.
\newblock U.S. Government Printing Office, Washington, DC, USA, tenth printing
  edition, 1972.

\bibitem{VerweyOverbeek-1948}
E.~J.~W. Verwey and J.~Th.~G. Overbeek.
\newblock {\em Theory of the Stability of Lyophobic Colloids}.
\newblock Elsevier, Amsterdam, 1948.

\end{thebibliography}
\end{document}